\documentclass[preprint]{elsarticle}

\usepackage{amsmath}
\usepackage[lofdepth,lotdepth]{subfig}
\usepackage{placeins}

\usepackage[final]{changes} 

\usepackage{lineno,hyperref}
\modulolinenumbers[5]
\usepackage{framed}
\usepackage{nomencl}
\makenomenclature

\usepackage{etoolbox}
\renewcommand\nomgroup[1]{%
	\item[\bfseries
	\ifstrequal{#1}{P}{Parameters of interest}{%
		\ifstrequal{#1}{V}{Time dependent variables}{%
		\ifstrequal{#1}{A}{Abbreviations}{}}}%
	]}

\journal{Applied Energy}

\begin{document}

\begin{frontmatter}

\title{Influence of weather natural variability on the thermal characterisation of a building envelope}


\author[mymainaddress]{Sarah Juricic\corref{mycorrespondingauthor}}
\cortext[mycorrespondingauthor]{Corresponding author}
\ead{sarah.juricic@gmail.com}

\author[mymainaddress]{Jeanne Goffart}
\author[mymainaddress]{Simon Rouchier}
\author[mysecondaryaddress]{Aurélie Foucquier}
\author[mymainaddress]{Nicolas Cellier}
\author[mymainaddress]{Gilles Fraisse}

\address[mymainaddress]{Univ. Savoie Mont-Blanc, CNRS, LOCIE, 73000 Chambéry, France}
\address[mysecondaryaddress]{Univ Grenoble Alpes, CEA, LITEN, INES, 38000 Grenoble, France}

\begin{abstract}
	\begin{sloppypar}
		\replaced{The thermal characterisation of a building envelope is usually best performed from on site measurements with controlled heating power setpoints. Occupant-friendly measurement conditions provide on the contrary less informative data. Notwithstanding occupancy, the boundary conditions alone contribute to a greater extent to the energy balance. Non intrusive conditions question therefore the repeatability and relevance of such experiment.}{The thermal performance assessment of a building envelope from in-situ measurements is usually best performed with a controlled heating power set point. Such experiment induces high indoor temperature and is incompatible with occupancy. Keeping non intrusive, occupant-friendly conditions provides less informative data compared to controlled and optimized indoor conditions. Notwithstanding occupancy, the boundary conditions alone contribute indeed to a greater extent to the energy balance in a non intrusive framework. This might affect uncertainties, bias the estimations of the thermal characteristics of interest and questions therefore the repeatability and relevance of such experiment.}
		
		\replaced{This paper proposes an original numerical methodology to assess the repeatability and accuracy of the estimation of an envelope's overall thermal resistance under variable weather conditions. A comprehensive building energy model serves as reference model to produce multiple synthetic datasets. Each is run with a different weather dataset from a single location and serves for the calibration of an appropriate model, which provides a thermal resistance estimate. The estimate’s accuracy is then assessed in the light of the particular weather conditions that served for data generation. The originality also lies in the use of stochastically generated weather datasets to perform an uncertainty and global sensitivity analysis of all estimates with respect to 6 weather variables.}{This paper proposes an original numerical methodology to assess the repeatability and accuracy of the estimation of an envelope's overall thermal resistance ($R_{eq}$) under variable weather conditions. The methodology is numerically based. A comprehensive building dynamic energy model serves as reference model to produce multiple synthetic datasets. Each is run with a different weather dataset from a single location and serves as calibration data for the calibration of a model $\mathcal{M}$, which in turn provides an estimate of $R_{eq}$. Accuracy of each estimation can then be assessed in the light of the particular weather conditions that served for data generation. The originality also lies in the use of stochastically generated weather datasets to perform an uncertainty and global sensitivity analysis of the estimates' accuracy with respect to 6 weather variables.}
		
		\replaced{The methodology is applied on simulated data from a one-storey house case study serving as reference model. The thermal resistance estimations are inferred from calibrated stochastic RC models. It is found that 11 days are necessary to achieve robust estimations. The large air change rate in the case study explains why the outdoor temperature and the wind speed are found highly influential.}{The methodology is applied on simulated data from a multi-zone one-storey house case study serving as reference model. The estimations of $R_{eq}$ are inferred from the calibration of stochastic RC models. It is found that a minimal observation duration of 11 days is necessary. The remaining variability is found to be exclusively influenced by the outdoor temperature and wind speed variability. This is due to the large air change rate in the reference model, inducing temperature and wind speed dependent ventilative losses.}
	\end{sloppypar}
\end{abstract}

\begin{highlights}
	\item \added{Thermal characterisation of a building envelope influenced by weather conditions}

	\item \added{Original methodology assesses minimal measurement duration for robust estimation}

	\item \added{Thermal characterisation by RC models is accurate and faster than steady state methods}

	\item \added{Necessary 11 day measurements for robust estimation sets new benchmark value in field}
	
	\item \added{Estimations highly influenced by outdoor temperature and wind speed in case study}
%
%
\end{highlights}


\end{frontmatter}


\begin{table*}[ht!]
	\begin{framed}
		\nomenclature[P,01]{$HTC$}{Heat Transfer Coefficient: overal heat transfer of the building envelope towards exterior as defined in \cite{ISO13789} (W/K)}
		\nomenclature[P,02]{$R_{eq}$}{Equivalent overal thermal resistance of the envelope, inverse of HTC (K/W)}
		\nomenclature[P,03]{$R^*_{eq}$}{Target thermal resistance: overal theoretical thermal resistance of the reference model (K/W)}
		\nomenclature[P,04]{$U$-value}{Thermal transmittance of a wall ($W/m^2K$)}
		\nomenclature[P,05]{$C_w$, $C_i$, $R_o$, $R_i$, $A_w$}{Parameters of the $2^{nd}$ order model $T_wT_i$~$R_oR_i$~$A_w$}
		
		\nomenclature[V]{$P_{heating}$ or $P_h$}{Heating power delivered in the building (W)}
		\nomenclature[V]{$T_{in}$}{Indoor temperature ($^{\circ}C$)}
		\nomenclature[V]{$T_{out}$}{Outdoor temperature ($^{\circ}C$)}
		\nomenclature[V]{$I_{sol}$}{Solar global horizontal irradiation ($W/m^2$)}
		\nomenclature[V]{$T_w$}{Unmeasured temperature of the building envelope ($^{\circ}C$)}
		\nomenclature[V]{$Q^{in}_{storage}$ and $Q^{out}_{storage}$}{Heat transfered in or out the building envelope ($W$)}
		\nomenclature[V]{$Q_{sun}$}{Heat gained indoors by solar irradiation ($W$)}
		\nomenclature[V]{$Q_{ventilation}$}{Heat transfers by air change in the building ($W$)}
		\nomenclature[V]{$Q_{ground}$}{Heat transfers from the indoors towards the ground ($W$)}
		
		\nomenclature[A]{ARX}{Auto Regressive Models with eXternal inputs}
		\nomenclature[A]{TMY}{Typical Meteorological Year}
		\added{\nomenclature[A]{ML}{Maximum-Likelihood}}
		\added{\nomenclature[A]{IWEC}{International Weather for Energy Calculations}}
		
		\printnomenclature
	\end{framed}
\end{table*}

\newpage

\listofchanges

\section{Introduction}
\label{S:Intro}


\subsection*{Background and motivation}

\begin{sloppypar}
	The renovation of buildings is a growing concern for the reduction of their global energy consumption \added[comment={additionnal background on energy conservation measures policies}]{as stated by the European Commission in its strategic long-term policy for a climate neutral enonomy \cite{Commission2018}. As underlined in the contribution \textit{Buildings} to the Fifth Assessment Report of the Intergovernmental Panel on Climate Change \cite{Lucon2014}}\replaced{, t}{. T}here is therefore a need for accurate estimations of the thermal performance of building envelopes in order to drive relevant energy conservation measures. \added{Heo et al \cite{Heo2013} show for example how the estimation of the actual performance of a building benefits a retrofit analysis even under uncertainty. }When known, the thermal performance of an investigated building serves indeed the energy retrofit decision plan \deleted{\cite{Heo2013}} by accurately reflecting on the possible energy gains.
	
	On site monitoring has shown to be a promising lead to perform an accurate thermal characterisation of the envelope. In particular, the estimation of the Heat Transfer Coefficient ($HTC$) or its inverse the overall thermal resistance ($R_{eq}$) has been studied as a result of controlled experiments during one or several days \cite{Thebault2018, Ghiaus2019, Bacher2011, Jimenez2008}, up to several weeks as in \cite{Jack2018,Alzetto2018, Brastein2018, Brastein2019}, mostly in unoccupied buildings and controlled indoor conditions. Uncertainty of the estimation of the thermal performance from on site measurements is indeed usually reduced by means of an optimally designed heating control in the building.
	
	In the case where controlled measurement conditions cannot be achieved, such as in continuously occupied buildings (hospitals, elderly homes,...), thermal characterisation may only be estimated from non intrusive \replaced{monitoring and a reduced number of sensors}{and reduced number of sensors and monitoring}. This implies that the sensors should not alter the building envelope and that occupants should not be disturbed by the monitoring equipment. A non intrusive approach then relies on possibly very few sensors in non optimally controlled and possibly only partially known operating conditions. On the other hand, relying on fewer sensors may positively contribute to providing a test easier to implement and less costly, if it is found to be feasible and accurate.
	
	Feasibility and accuracy are indeed questioned by the uncontrolled nature of such experiment. The building envelope energy balance as shown in Equation \ref{eq:building envelope energy balance} in an uncontrolled experiment shows different dynamics than in controlled experiments. The heating power delivered indoors $P_{heating}$ is less informative than when it is optimally designed. The outdoor boundary conditions by their contributions through $T_{out}$, $Q_{ventilation}$ and $Q_{sun}$ have therefore a proportionally larger influence on the energy balance. This will particularly be the case when shorter datasets are used to infer the $HTC$ or $R_{eq}$, whereas longer experiments provide de facto a wider natural variability of the boundary conditions which should flatten the effect of uncommon punctual outdoor conditions.
	
	\begin{equation}
	\centering
	\begin{array}{r c l}
	HTC \cdot (T_{in}(t) - T_{out}(t)) + Q^{in}_{storage}(t) + Q^{out}_{storage}(t) & &
	\\
	+ Q_{ventilation}(t) + Q_{ground}(t) - P_{heating}(t) - Q_{sun}(t) & = & 0
	\end{array}
	\label{eq:building envelope energy balance}
	\end{equation}
	
	In addition, the outdoor weather conditions have correlated frequencies, which are themselves correlated to common indoor temperature setpoint schedules. This may impede the accuracy of the estimation, let alone the identifiability of the heat transfer coefficients.
	
	For these reasons, the estimation of the $HTC$ or $R_{eq}$ coefficients in non intrusive conditions may be considerably influenced by the boundary conditions. The amount of data needed to secure a robust and accurate $HTC$ or $R_{eq}$ estimation could be significantly larger than what is usually considered in literature for tests in a controlled framework. The risk with too short datasets is to obtain significantly different results following identical procedures but carried out a day, a week or a year later. Non intrusive conditions are indeed poorly informative for the estimation process: the non intrusive framework questions the uncertainty level and therefore the repeatability of such estimation.
	
\end{sloppypar}
	
\subsection*{Estimation of the U-value, $HTC$ or $R_{eq}$ coefficients in non intrusive conditions: models, experiment duration and influence of weather conditions in the existing literature}

\begin{sloppypar}
	Previous work on thermal performance estimation of the envelope in a non intrusive framework has focused on the thermal characterisation of either a single wall performance or the entire building envelope. To the extent of our knowledge, how variability of the weather conditions influences the accuracy of the estimation has never been extensively developed. However, the following papers address the issues of convergence of the estimation and measurement duration with, for some, insight on the influence of outdoor conditions on the quality of the results.
	
	At wall scale, Rasooli and Itard \cite{Rasooli2018} show from numerically generated data how solar irradiation significantly defers stability and convergence of the estimation of the conductive thermal resistance of a wall $R_c$ using heat flow meters following the ISO 9869 standard \cite{ISO9869}. Simulated data with and without solar irradiation showed indeed significantly different convergence towards the final estimation value. The authors suggest using flow meters on both sides of the walls to secure a robust and faster estimation of $R_c$.
	
	Petojevic et al \cite{Petojevic2018} propose an innovative method to exploit non intrusive data from heat flux and temperature meters to determine dynamic thermal characteristics of a wall. The use of 12.5 days data, although not justified, met acceptable accuracy on the results.
	
	Gaspar et al \cite{Gaspar2018} precisely study the minimal duration of a heat flux meter test for estimating the $U$-value of a wall by the average and dynamic methods of the ISO 9869 standard \cite{ISO9869}. The authors compare the stability of the estimation to the criteria given by the ISO 9869 standard \cite{ISO9869}, in which is given that three conditions must be met simultaneously to end the test:
	
	\begin{itemize}
		\item The first condition is that the test must last 72 h or longer,
		\item The second condition is that the $U$-value obtained at the end of the test must not deviate more than 5~\% from the value obtained 24 h earlier,
		\item The third condition is that the $U$-value obtained from the first N days and from the last N days must not deviate more than 5~\%, with $N=2/3\times$~total duration.
	\end{itemize}
	
	Gaspar et al \cite{Gaspar2018} conclude that these conditions provide trustworthy estimations and secure convergence towards the final value within 4 to 5 days for both the average and dynamic methods described by the ISO 9869 standard \cite{ISO9869}. Influence of the weather variability on the convergence rate has not been investigated. Higher actual transmittance of the façades were found to be the main reason for slower convergence.
	
	At wall scale again, Gori and Elwell \cite{Gori2018a} as well as Gori et al \cite{Gori2018b} exploit heat flux measurements with RC models and introduce the idea of stabilisation of the estimation: from short datasets, the estimates suffer from the prominent noise in the data. As the dataset grows, the values stabilise towards a final value. Applying the criteria of the ISO 9869 standard \cite{ISO9869}, they found that up to 10 days were necessary to reach stabilisation in autumn and winter season whereas longer periods were necessary in warmer seasons. The minimum length tested was 3 days, as demanded by the ISO 9869 standard \cite{ISO9869}, but authors found that shorter datasets sufficed in some cases with the use of a RC model, implying that the three conditions of the ISO 9869 standard \cite{ISO9869} might be too conservative when applied to other methods.
	
	At wall scale too, Rodler et al \cite{Rodler2019} compared a dynamic model calibrated by Bayesian inference to the average and dynamic methods described in the ISO 9869 standard \cite{ISO9869} and found that the temperature difference was more determinant than the length of the dataset, therefore confirming the significant role of boundary conditions in uncontrolled experiments.
	
	Deconinck and Roels \cite{Deconinck2017} applied dynamic grey box modelling in a non intrusive framework to assess the thermal performance of a single wall based on heat flux measurements. The authors used two different data subsets of 10 days in winter (steady indoor temperature assumed at 20~$^{\circ}C$) and 9 days in summer (free floating indoor and outdoor temperatures). They found that winter conditions with constant indoor temperatures were not appropriate to identify the parameters of interest, considering that temperatures are the main variables of the differential equations used for the exploitation of the data. Summer free floating conditions were then found to be more informative and led to identifiable and interpretable parameters. Let us remind here that identifiability relates to the unicity of the parameter estimation and interpretability to the ability to give the estimation a physical meaning. Both may be confounded if the model characterizes perfectly the system.
	
	At building scale, Redyy et al \cite{Reddy1999} point out the issue of data informativeness as well as influence of weather conditions when assessing the overall heat loss and overall ventilation rate of a large commercial building. To perform the assessment, non intrusive measurements are averaged and exploited by a steady-state equation. They found that daily averaged data over a year combined with a multi-step regression technique, where multiple regressions are performed one after the other to estimate parameters one by one, achieved the best results. Parameter identification over a single season was less accurate: in winter and summer seasons, the combined variability of the outdoor temperature and the relative humidity was narrower than during the spring season. Large variability of these two weather variables yielded less correlated parameters and more accurate overall parameter identification.
	
	More recently, Senave et al \cite{Senave2019} studied the physical interpretation of auto-regressive models with exogenous inputs (ARX), aiming for the estimation of the HTC via on-board monitoring, i.e. in a non intrusive measurement framework. Four different indoor temperature scenarii were twice tested through 20 days of synthetic data: once for calibration and once for validation. The building model is a single-zone opaque box and the study focused on the estimation of the HTC in case of heat losses to the ground. There is no mention of the influence of the measurement duration on the results. In Senave et al \cite{Senave2020}, ARX models and linear regressions were compared to exploit data collected from again non intrusive experiments on datasets of 26 weeks.
	
	Grey-box models such as stochastic RC models have not been used to exploit data at building scale, but showed promising results at wall scale as in \cite{Deconinck2017} or \cite{Gori2018b} with shorter datasets than the average method of the ISO 9869 standard \cite{ISO9869}. Stochastic RC models have also been used in controlled experiments and provide satisfactory $HTC$ estimation from short datasets \cite{Thebault2018}.
	
	In conclusion, minimal duration for heat flow measurements to infer the actual thermal transmittance of a single wall has been addressed in literature and relevance of the ISO 9869 standard \cite{ISO9869} criteria discussed. At wall scale, accurate estimations of U-values may be performed within a week, sometimes quicker in the case of low transmittance façades. Solar irradiation seems to be influential on the convergence rate.
	
	For a better representation of the overall thermal performance of the envelope, some methods aim to the $HTC$ estimation at building scale in non intrusive experiments. Literature however is scarce and the assessed methods use much larger datasets, which increases the variability in the data and therefore reduces the influence of boundary conditions on the quality of the result. It has also been suggested that large variability of the outdoor temperature and relative humidity led to a better identification.
\end{sloppypar}

\subsection*{Objectives of this work}

\begin{sloppypar}
	Accurate estimation of the $HTC$ or $R_{eq}$ coefficients at building scale in non intrusive experiments would certainly be beneficial as guidance for relevant energy conservation measures. Although the existing literature suggests that such estimation is feasible from datasets of several weeks, shorter experiments are desirable to decrease immobilisation of the measurement devices and cost of the procedure. Feasibility and accuracy of the $HTC$ or $R_{eq}$ identification from shorter datasets is yet questioned under naturally variable weather conditions given their larger contribution to the building envelope energy balance.
	
	This paper therefore intends to assess the feasibility, i.e. the accuracy and repeatability of the $R_{eq}$ estimation from non intrusive experiments. The underlying hypothesis is that there is a minimal measurement duration after which the $R_{eq}$ estimation keeps steady, regardless of the boundary conditions. Provided that the measurement duration is sufficient, a $R_{eq}$ estimation could then start at any time and on any day under any usual weather conditions, the estimation should remain robust. Accuracy and feasibility of the estimations are therefore assessed in the light of the natural and local variability of weather conditions.
	
	This paper proposes to perform the estimation with stochastic RC models in the hope of achieving reasonably fast estimations of the overall thermal resistance of the envelope $R_{eq}$, as was suggested by the literature review.
	
	The investigation of the feasibility of such estimation is based on an original numerically based methodology. Indeed, simulated datasets from a reference model are used to calibrate stochastic RC model and infer an estimation of $R_{eq}$. The characteristics of the reference model and the case study that served for data generation are presented in Section \ref{S:Ref model}. As detailed in Section \ref{S:Methodo}, weather natural variability is introduced by the use of multiple weather datasets for the simulations of the reference model. In particular, the weather datasets used are stochastically generated as to explore to full width of natural variability that can be expected in a typical January. The repeatability of the estimation of the thermal resistance is then discussed in Section \ref{S:variability}. Section \ref{S:influential} examines the influence of the 6 stochastically generated weather variables on the thermal resistance estimates. Section \ref{S:Discussion} finally puts the results in perspective with the previous literature review and discusses the relevance of using the ISO 9869 standard \cite{ISO9869} criteria to assess the convergence of an estimation.
\end{sloppypar}

\section{Reference model and description of the case study}
\label{S:Ref model}

The methodology developed and applied in this paper is based on a numerical reference building energy model. The modelling choices, the case study and its thermal characteristics are presented in this section.

\subsection{Relevant modelling choices for providing synthetic data}

Various choices can be made for the reference model to account for heat and mass transfer modelling, solar irradiation, etc. The choices made for the reference model are therefore driven by the purpose of this study, i.e. thermal behaviour, but also on the need of reasonable simulation duration as many simulations are planned for the $R_{eq}$ estimation assessment.

Among the different algorithms EnergyPlus has implemented for heat and/or moisture transfers in the building, the \textit{Conduction Transfer Function} seems to be an appropriate option: it is a heat only algorithm and does not account for moisture storage and diffusion. Using heat and moisture transfer algorithms will only slow the simulation time without adding significant
improvements in the assessment methodology. The \textit{Conduction Transfer Function} algorithm is fast as it relies on a state space representation with the finite difference wall temperatures as variables \cite{Seem1989}. From the state space representation, it is possible to formulate the model output as a direct function of the input, without calculating storage and temperatures at the discretization nodes in the wall.

As for heat transfers through ventilation and infiltration, EnergyPlus has two possible options: the \textit{DesignFlowRate} module and the \textit{Airflow Network} model. The \textit{Airflow Network} model, based on pressure and airflow calculations with temperature and humidity calculations, has however been developed to simulate with accuracy air distribution systems and its performance, such as supply and return duct leaks, multi-zone airflows driven by outdoor wind and mechanical ventilation. Although such detail is without question physically more accurate, it is also much more computation consuming. The \textit{DesignFlowRate} module, concededly simple, has been rather chosen for the reference model. Infiltration and ventilation flow rates are accounted for by the same module, but in separate inputs such as to enable different values. It relies
on equation \ref{eq:airflow rate EP} to calculate the airflow rate at each time step:

\begin{equation}
	Q(t) = V_{design} \cdot F_{schedule} \cdot (A + B \cdot |T_{zone} - T_{odb}|  + C \cdot WindSpeed + D \cdot WindSpeed^2 )
	\label{eq:airflow rate EP}
\end{equation}

\begin{tabular}{r l}
	where & $V_{design}$ the air flow rate ($m^3/s$) \\
	& $F_{design}$ an optional schedule that can vary over time,\\
	& $A$, $B$, $C$ and $D$ coefficients between 0 and 1,\\
	& $T_{zone}$ the zone indoor air temperature ($^\circ C$),\\
	& $T_{odb}$ the outdoor dry bulb temperature ($^\circ C$).
\end{tabular}

$A$, $B$, $C$ and $D$ are fixed identically as the default BLAST (EnergyPlus predecessor) \cite{BigLadderSoftwareVentilation}: \[A=0.606, B=0.03636, C=0.1177, D=0\]

\begin{sloppypar}
	Solar irradiation plays a large role in the building energy balance. In particular as the window blinds are maintained open in the simulations, solar irradiation entering the building through the windows come from multiple sources: direct beams with time-dependent values for each wall, diffuse irradiation by environment, reflections of direct irradiation on environment as well as indoor diffuse reflections. To account for such details, the \textit{Full Interior And Exterior With Reflections} EnergyPlus module is used.
	
	The EnergyPlus simulation run period extends from approximately November $1^{st}$ to March $31^{st}$ to cover a winter season. Furthermore, accurate estimations are expected to be larger in the coldest months, from December to February. This is due to the larger temperature difference between indoors and outdoors and lower solar gains. This implies that the energy balance during winter days is more sensitive to the thermal performance of the building envelope. It is however important to start the simulation period earlier as to avoid any impact of the warm up runs performed by EnergyPlus. Indeed, the software runs multiple times the same first day, beginning at an indoor temperature of 23 $^{\circ}$ C until convergence of the indoor conditions is met. The warm up allows credible initial conditions for the rest of the simulation. But, as the first weather day of the simulation period is random, it might put a particular weight on a possibly unusual cold or warm day, thus misleading the first hours and days of simulation, depending on the thermal mass of the building. It is then safe to start early in the winter period and then discard the first 15 days of simulation. In the end, the simulation run period starts on November $15^{th}$and finishes on February $15^{th}$.
	
	As for the simulation time step, it should not be larger than around 15 minutes, and preferably lower in case this reference model has lower characteristic times. Time steps larger than 15 minutes could hide aliasing in the data: short but influential phenomena are not seen. Aliasing in data then leads to potentially dramatically wrong estimations \cite{Madsen2007}. In this case study, it is fixed at 10 minutes.
	
	One of the identified pitfalls for generalisable model assessment framework is the fact that the simulation output are "ideal" measurements: the simulation output is deterministic. There are no systematic error and no random measurement error. To maintain the focus on the influence of weather variability on thermal characterisation, systematic errors will not be added to the outputs. Measurement random error on the other hand are a non negligible part of the issue of solving inverse problems \cite{Maillet2011}. In agreement to \cite{Leroy2010}, \cite{Sengupta2015} and \cite{Stoffel2000}, white noise is added to the following simulation outputs:
\end{sloppypar}

\begin{itemize}
	\item temperatures: addition of a normal noise $\mathcal{N}(0,~0.2~^\circ C)$,
	\item heating power: addition of a normal noise $\mathcal{N}(0,~20.0~W)$,
	\item solar irradiation: addition of a normal noise $\mathcal{N}(0,~5.0~W/m^2)$.
\end{itemize}

\subsection{The case study}

\begin{figure}
	\centering
	\includegraphics[width=0.6\textwidth]{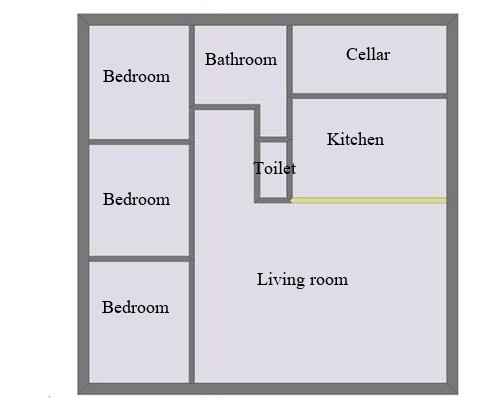}
	\caption{Floor plan of the one storey house serving as reference model}
	\label{fig:ref_building}
\end{figure}

The case study with which the methodology is applied in this paper is a multi-zone building of a one-storey house, as shown in Figure \ref{fig:ref_building}. The heated space is about 98~m$^2$ and has a total volume around 250~m$^3$. The building is equipped with convective heaters. The air change rate is 1.0 volume per hour.

It has unheated and unventilated crawlspace and attics. Heat losses towards the crawlspace may be considered insignificant, as the insulation layer under the concrete slab of the ground floor has been set at 30~cm. As for the rest of the building envelope, exterior walls are constituted of a 20~cm brick wall, with 10~cm insulation and 1~cm plaster on the interior side whereas the attics and the indoor space are separated by a 1~cm plaster and 30~cm insulation. All windows, frames included, have U-values between 1.3 and 1.6~$W/m^2K$. Total window surfaces add up to 15.9~m$^2$, among which 0.6~m$^2$ north, 5.4~m$^2$ east, 6.7~m$^2$ south and 3.2~m$^2$ west. The shading facilities are not activated and allow therefore solar gains. \added{South and East facing facades have a shading overhang, designed to avoid solar irradiation in summer. In winter conditions, most of the irradiation enters the envelope. }Table \ref{tab:proprietes case study} then summarizes the thermal properties of interest in this case study and Figure \ref{fig:ref_building} shows how the building is configured.

\begin{table}[h!]
	\centering
	\begin{tabular}{c|c|}
		\cline{1-2}
		\multicolumn{1}{|c|}{Vertical insulation thickness}              & 10~cm         \\ \hline
		\multicolumn{1}{|c|}{Attic insulation thickness}          & 30~cm         \\ \hline
		\multicolumn{1}{|c|}{Ground floor slab insulation thickness} & 30~cm         \\ \hline
		\multicolumn{1}{|c|}{Air change rate} & 1.0 h$^{-1}$         \\ \hline
	\end{tabular}
	\caption{Thermal characteristics of the case study used in this application}
	\label{tab:proprietes case study}
\end{table}

\begin{figure}
	\centering
	\includegraphics[width=\textwidth]{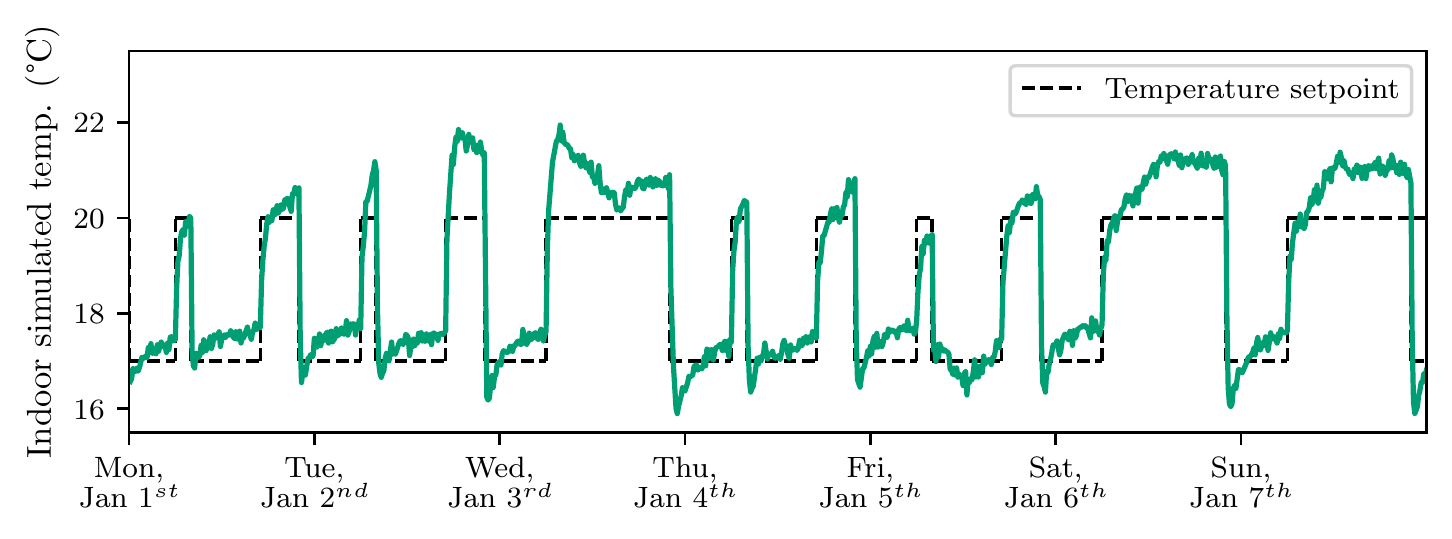}
	\caption{Illustration of \replaced{1 week simulated indoor temperature with added noise: Wednesday and week-ends have different temperature setpoints.}{3 days simulated data extraction: temperatures and solar irradiation with added noise}}
	\label{fig:temp_setpoint}
\end{figure}

The indoor temperature set point schedule is designed to mimic occupant-friendly conditions to meet the objective of studying how poorly informative data influences interpretability. Seen that dynamic models such as RC models cannot adequately learn from data in close to steady state conditions, a realistic temperature setback is therefore scheduled and follows a usual occupant related schedule. The indoor temperature schedule is set to reach 20~$^{\circ}$C in the morning and in the evening for workdays, and all day long during week-ends and on Wednesdays. The rest of the time, the temperature is scheduled to keep at 17~$^{\circ}$C. Figure \ref{fig:temp_setpoint} illustrates a week of simulated indoor temperature with such schedule.\added{ Noteworthy in this figure is that, as the indoor temperature setpoint is based on the operative temperature, there are slight differences between the simulated indoor air temperature and the temperature setpoint.}

\FloatBarrier

\subsection{Thermal performance of the case study}
\label{S:calcul Req}

As mentioned before, basing the procedure on simulated data offers the advantage that a theoretical thermal resistance of the envelope is known. In our case, it is determined by a simulation run with constant boundary conditions: no solar irradiation, constant indoor and outdoor temperatures. Wind speed however is non null and is kept at the values of the $TMY$ file of Geneva. Removing the wind speed would have decreased the overall heat loss coefficient, by diminishing the heat losses by ventilation as ventilation has been modelled to be wind speed dependent. Since in the synthetic experiment the ventilation is not stopped, the $R_{eq}$ estimations would not have converged towards a target value calculated without wind.

With these steady-state boundary conditions, the dynamic terms $Q^{in}_{storage}$ and $Q^{out}_{storage}$ in the energy balance from Equation~\ref{eq:building envelope energy balance} become negligible and Equation~\ref{eq:linreg} therefore perfectly describes the linear relationship between the heating power and the indoor-outdoor temperature difference:

\begin{equation}
\mathit{P_\mathit{heating}} =  HTC \times (T_\mathit{in} - T_\mathit{out}) = 1 / R^*_{eq} \times (T_\mathit{in} - T_\mathit{out})
\label{eq:linreg}
\end{equation}

A least square regression is performed on daily averaged data from January, February and March (92 days) and gives $R^*_{eq} = 5.19 \times 10^{-3}$~$K/W = 5.19~K/kW$ with a Pearson ($R^{2}$) coefficient of the linear regression of 0.999, showing an excellent fit. This value is from now on called \textbf{\textit{target}} $R^*_{eq}$. Means of comparison of an estimated $R_{eq}$ to the target value $R^*_{eq}$ is described in section \ref{S:interpret indic}.

\FloatBarrier

\section{Accuracy assessment methodology}
\label{S:Methodo}

\subsection{Overview of the applied methodology}

\begin{sloppypar}
	To assess the accuracy and repeatability of the thermal resistance $R_{eq}$ estimation of a building envelope while accounting for weather natural variability, the general idea is to use multiple datasets, each in different yet coherent weather conditions. A model $\mathcal{M}$ is calibrated from each dataset, providing as many estimation of its parameters $\theta$ as there are datasets. From the hence estimated parameters $\theta$, an estimation of $R_{eq}$ is inferred. Measurement duration and variability in the weather conditions themselves will then influence the accuracy and uncertainty of the estimations. In principle, for a certain minimal measurement duration, the final estimates are expected to show robustness and to remain significantly similar regardless of the natural variability in weather conditions.

	The essence of the methodology is therefore to use multiple datasets, each in different weather conditions. Datasets from actual on-site measurements during an entire heating season, or better during several heating seasons would concededly ensure strong realism. However, the use of real datasets may only deliver an incomplete view of the issue and be difficult to analyse. Indeed, as the initial conditions and the thermal state of the envelope would be different for each dataset given the previous days, analysing the variability of the $R_{eq}$ estimations could not only be attributed to weather variability.
	
	To avoid this pitfall, the proposed original methodology relies on a fully numerical 4 step procedure, as illustrated by Figure \ref{fig:num_proc}:

\begin{itemize}
	\item a computer based model serving as \textit{reference model} is implemented in a program for dynamic thermal simulations (here, Energy Plus). Given that the reference model is purely numerical, an analytical value of target $R^*_{eq}$ can be calculated. The reference model has been detailed in section \ref{S:Ref model} and in particular the calculation of target value $R^*_{eq}$ in section \ref{S:calcul Req}.
	
	\item Simulation and output processing (\textbf{step I} part 1): a dynamical simulation of the reference model is run under known boundary weather conditions. The choice of a set of synthetic weather datasets to perform the simulation is detailed in section \ref{S:Weather}. The simulation output provides synthetic data of the resulting indoor conditions.  White noise is added to mimic actual measurements of indoor and outdoor variables. Step (I) is repeated $n$ times, each time with a different weather dataset. Coming section \ref{S:Weather} further details what weather data is used to perform the energy simulations.
	
	\item Data subset extraction (\textbf{step I} part 2) Seven subsets of data are extracted from each simulated dataset, all starting on January $2^{nd}$ with growing lengths from 2 days to 25 days. Simulation parameterization and data selection is described in section \ref{S:Step I}.
	
	\begin{figure}
		\centering
		\includegraphics{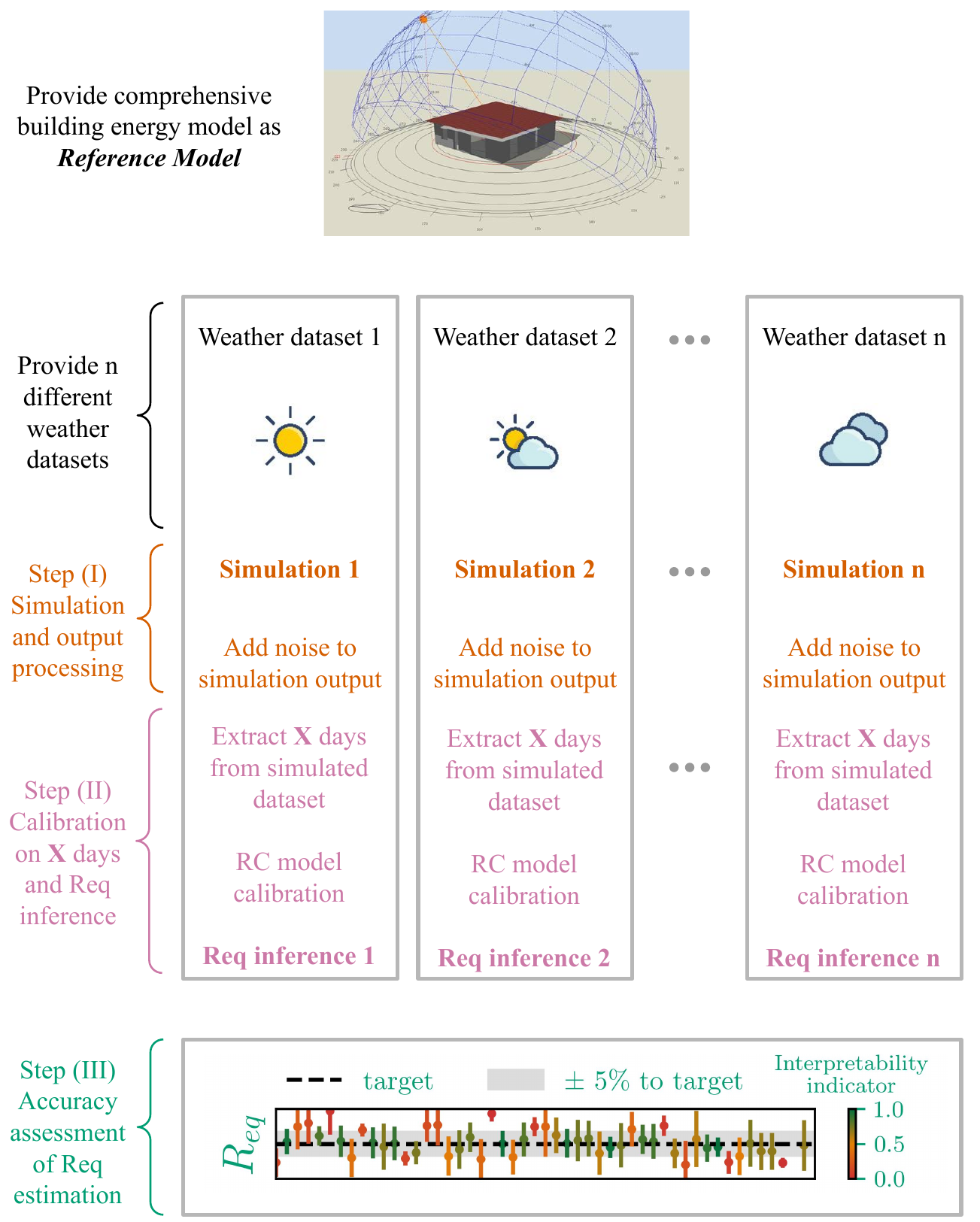}
		\caption{Methodology overview to assess the influence of weather variability over accuracy of a $R_{eq}$ estimation. In the end, each estimation is compared to the target value by means of a novel interpretability indicator.}
		\label{fig:num_proc}
	\end{figure}
	
	\item Calibration (\textbf{step II}): Model $\mathcal{M}$ under study, chosen to provide an estimate of $R_{eq}$, is calibrated on each subset of the simulation output data. Model calibration is performed from a frequentist approach with a BFGS algorithm minimizing the negative log-likelihood. The calibration procedure is thoroughly described in \ref{S:model calibration}. A $R_{eq}$ estimation can be derived from the estimated parameters of each calibrated model $\mathcal{M}$. All in all, step (II), that comprises parameter estimation of model $\mathcal{M}$ and subsequent $R_{eq}$ inference, is done $7 \times n$ times. All steps from choice of model to $R_{eq}$ inference are detailed in section \ref{S:step II}.
	
	\item Accuracy assessment (\textbf{step III}): the estimated $R_{eq}$ is compared to the target value $R^*_{eq}$ of the reference model by means of a novel interpretability indicator, described in section \ref{S:interpret indic}, which reflects on both the uncertainty and the relative error to the target value. The variability and accuracy of the $R_{eq}$ estimation are assessed by studying the evolution with growing measurement duration of the total variance of the maximum-likelihood estimates as well as that of the interpretability indicator. 
\end{itemize}

\end{sloppypar}

\FloatBarrier

\subsection{Providing weather data to the simulation step (I)}
\label{S:Weather}

\subsubsection{Using actual weather data: a limited insight}
\label{S:actual weather}

\begin{figure}
	\centering
	\includegraphics[width=\linewidth]{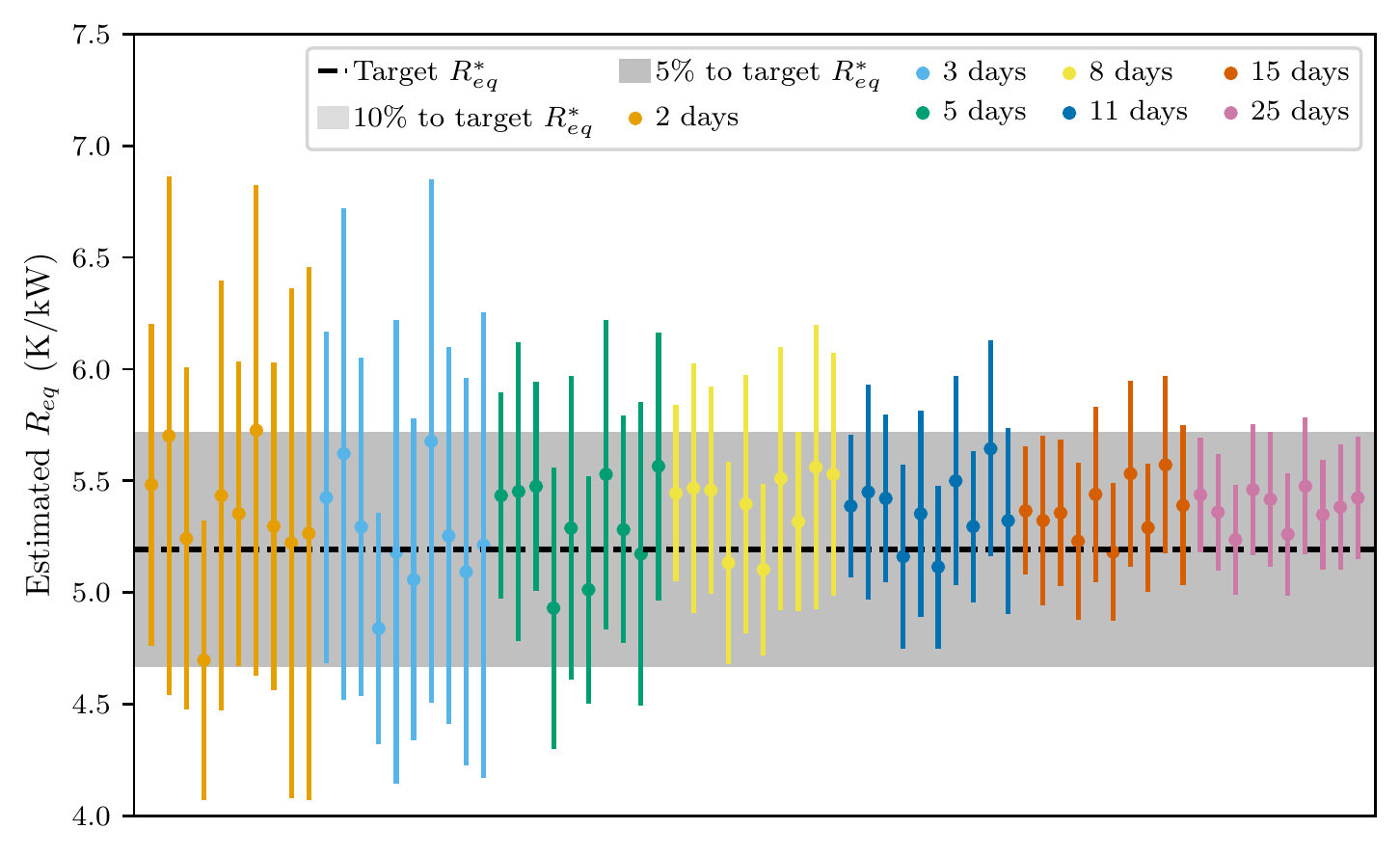}
	\caption{\added{Evolution of the variability of the estimation of $R_{eq}$ from 2, 3, 5, 8, 11, 15 and 25 day data subsets: data simulated with historical weather data of Geneva from 1990 to 1999} \deleted{Variability of the estimation of $R_{eq}$ with respect to real weather conditions: simulation and calibration from 11 weather data files and comparison of the estimation of $R_{eq}$ from 2, 3, 5, 8, 11 and 15 days data subsets}}
	\label{fig:rationale_geneva}
\end{figure}

As a first attempt to understand the influence of weather conditions, the reference model was run with 10 years of historical \added{winter }weather data in Geneva (Switzerland)\added{, from year 1990 to 1999 included}.\deleted[comment={estimation from TMY is unnecessary and not strictly speaking historical data}]{ An 11th simulation is performed with the \textit{TMY} weather file, representative of typical weather in Geneva.} From each simulation, subsets of data of variable durations are extracted: 2, 3, 5, 8, 11, 15 and 25 days. Model $T_wT_i$ $R_oR_i$ $A_w$ as in Equation \ref{eq:mod2R2C} is then calibrated. An estimation of $R_{eq}$ is inferred for each subset and shown in Figure \ref{fig:rationale_geneva}: each dot represents the maximum likelihood (ML-) estimate and the bars represent the confidence intervals. \added[comment={short explanation of what ML-estimation as full explanation of model calibration is given later}]{As will be specified in \ref{S:model calibration}, the ML-estimates are the most likely values for the parameters given the collected data.}

Preliminary results can be inferred from Figure \ref{fig:rationale_geneva}:

\begin{itemize}
	\item Short datasets provides $R_{eq}$ estimations with a large variability and high uncertainty. There seems indeed to be no agreement in the ML-estimates (visible as dots) and the confidence intervals are large. This suggests in particular that 2 or 3 days are insufficient for a robust estimation of $R_{eq}$.
	\item Variability of the ML-estimates decreases as the measurement duration increases. All ML-estimates converge within a 5~\% error band around the target $R^*_{eq}$ with 25 days datasets.
	\item regardless of the measurement duration, the significant variability of the ML-estimates can only be attributed to weather variability induced by the different weather datasets used for simulation. What particular conditions cause under or overestimation can however not be inferred from this first application.
\end{itemize}

\begin{sloppypar}
	These preliminary results suggest a decrease in variability with growing measurement duration. However, seeing that this first application is performed on only \replaced[comment={estimation from TMY is unnecessary and not strictly speaking historical data}]{10}{11} weather datasets, concluding on a minimal length would be statistically weak. In addition, the datasets do not allow to attribute the estimations variability to one or more particular weather variable.
\end{sloppypar}

\subsubsection{Stochastic weather data to perform global sensitivity analysis}
\label{S:stochastic weather}

\begin{sloppypar}
	To perform a more exhaustive assessment of the $R_{eq}$ estimations under variable weather conditions, the proposed methodology is now applied with a set of 2000 synthetic weather datasets with which a variance based sensitivity analysis is possible \cite{Goffart2017}.
	
	A total of 6 weather variables are stochastically generated to be representative of usual weather conditions in Geneva in winter, following the methodology described in \cite{Goffart2017}, as a time series constructed by a combination of statistical and deterministic features. The characteristics are extracted on the basis of the \added[comment={specification about the IWEC file that is used to generate the synthetic weather datasets}]{$IWEC$ weather data file (International Weather for Energy Calculations) \cite{EPlusIWEC} from Geneva. \textit{IWEC} files are built like }$TMY$ weather file\added{s} \cite{Pernigotto2014}\added{for locations outside the United States and Canada}. The \emph{TMY} file, standing for Typical Meteorological Years, is built by concatenation of typical months. Each month is chosen from 30 years actual data: each monthly dataset is weighted as a sum of 13 Finkelstein-Schafer statistics \cite{Finkelstein1971} from the temperature, wind and solar radiation data. In the end, the chosen monthly dataset is the one that shows statistics closest to mean, median of the 30 years data distribution, after having discarded years with exceptionally long periods of consecutive warm, cold or low radiation days. The stochastic generation \cite{Goffart2017} contains then as much variability than in the $TMY$ file: if the $TMY$ has for one particular variable a lower variability than the rest of the 30 years actual weather data, it will reflect in the synthetic data.
	
	From the $TMY$ file, Goffart et al \cite{Goffart2017} select 6 weather variables to stochastically generate 2000 weather files, the rest of the variables are left unchanged. The generated variables are exterior dry bulb temperature, relative humidity, direct normal solar irradiation, horizontal diffuse solar irradiation, wind speed and wind direction.
	
	Finally, the weather data is generated as to calculate sensitivity indices through a Sobol variance method able to cope with groups of time-dependent inputs, like here time dependency of each weather variable. Sensitivity indices by groups estimate the effect of the entire time series of the meteorological variable under study. The sensitivity indices are therefore scalars even though the variables are time series. The indices are calculated from two sets of 1000 samples, each sample of the first 1000 being defined by the characteristic features extracted from the $TMY$ file of each weather variable, the second 1000 samples being a rearrangement of the first.
	
	In this study, the output of interest for the sensitivity analysis is the $R_{eq}$ estimation and in particular the weather conditions leading to an increased or decreased estimation.
	
	In order to check the representativeness of the generated weather data, Figure \ref{fig:6_representativeness_stoch_data} compares the synthetic data to the actual historical data from Geneva and to the \textit{TMY} data. The figure shows the empirical cumulative distributions of the 6 weather variables for the month of January of the historical weather data in black thin lines and in orange the \textit{TMY} data. The grey areas represent the 50~\%, 75~\% and 95~\% quantiles of the synthetic data. The lines represent the cumulative distributions of all 6 weather variables. The higher the line, the lower the values of its time series. For example, the wind speed in the $TMY$ file, in orange, is lower than any other historical weather data which means that the $TMY$ file is in overall in January more windy than the 10 years of historical weather data.
\end{sloppypar}

From Figure \ref{fig:6_representativeness_stoch_data} can be inferred that:

\begin{figure}
	\centering
	\includegraphics[width=\textwidth]{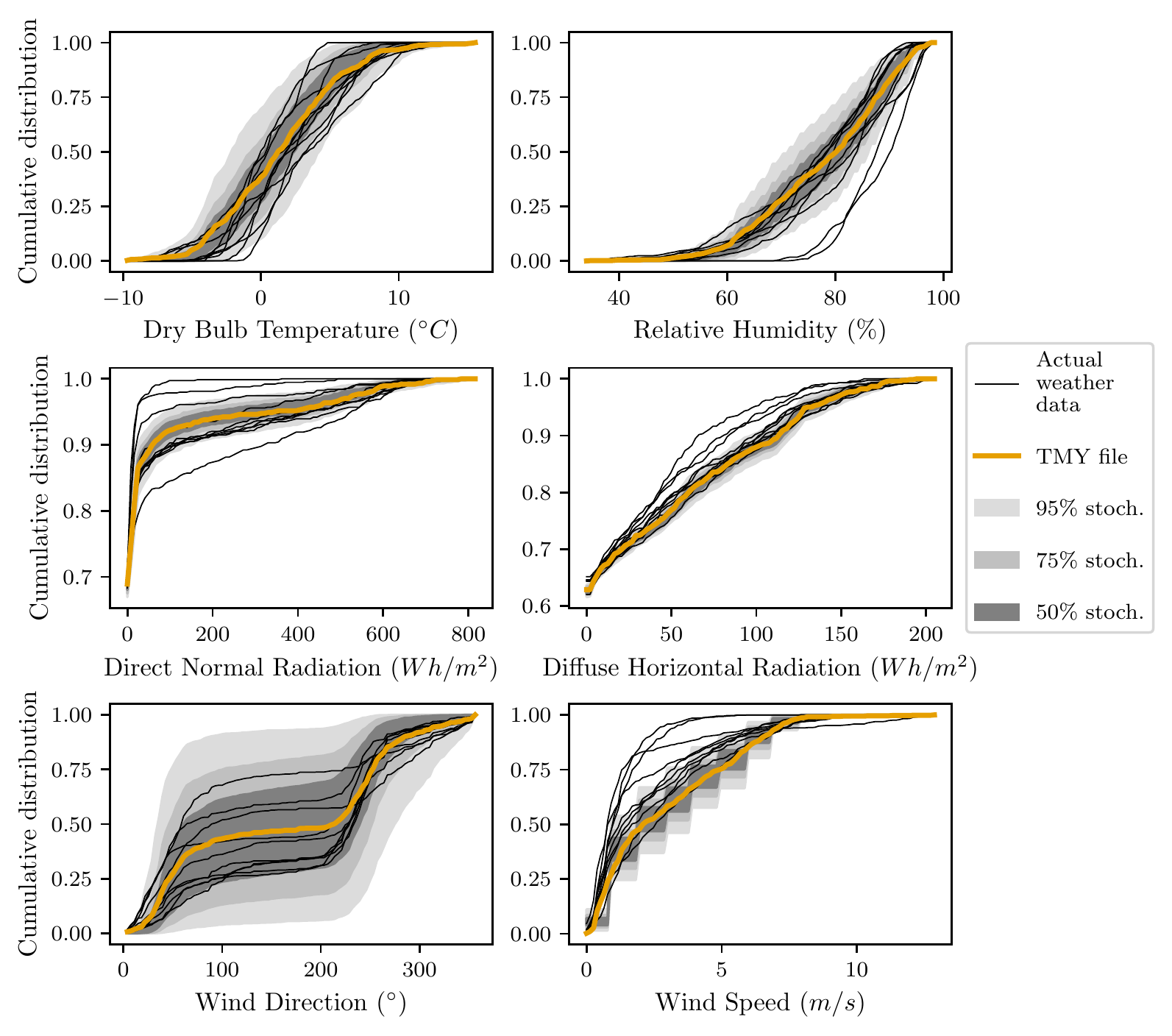}
	\caption{Representativeness of the stochastic weather data with respect to real weather data: the lines represent the cumulative distributions of all 6 weather variables. The generated outdoor temperature and direct normal irradiation data in grey is representative of historical data whereas wind speed is rather in the generated data than in the historical measurements in Geneva.}
	\label{fig:6_representativeness_stoch_data}
	\vspace{0pt}
\end{figure}

\begin{itemize}
	\item Synthetic outdoor dry bulb temperatures seem to be representative of the historical measurements. Synthetic wind direction is in good agreement with the historical measurements as well.
	\item Synthetic relative humidity seems to be lower than some of the historical measurements. The synthetic diffuse radiation on the contrary seems slightly overestimated, as does the wind speed.
	\item The generated direct normal radiation data does not cover a range as wide as the actual data: some of the real data may have much higher or lower direct radiation. This might have an impact on the following results and will be discussed later on.
\end{itemize}

\FloatBarrier
\subsection{Step (I): Simulation, output processing and data subset selection}
\label{S:Step I}

\subsubsection{Simulation}

For the purpose of the study, the thermal simulation of the multi-zone reference model needs to be performed such as to deliver energy consumption and temperature. Let us also shortly remind that the reference model is run on a winter season, from November 15th to February 15th. The time step of the output is set at 10 minutes in order to catch higher frequency phenomena and improve the accuracy of the estimation \cite{RamosRuiz2017}. Winter season secures outdoor temperatures below 15~$^{\circ}C$ which creates a significant temperature difference with indoors and enhances the practical identifiability of the parameters of interest.

\begin{figure}
	\centering
	\includegraphics[width=\textwidth]{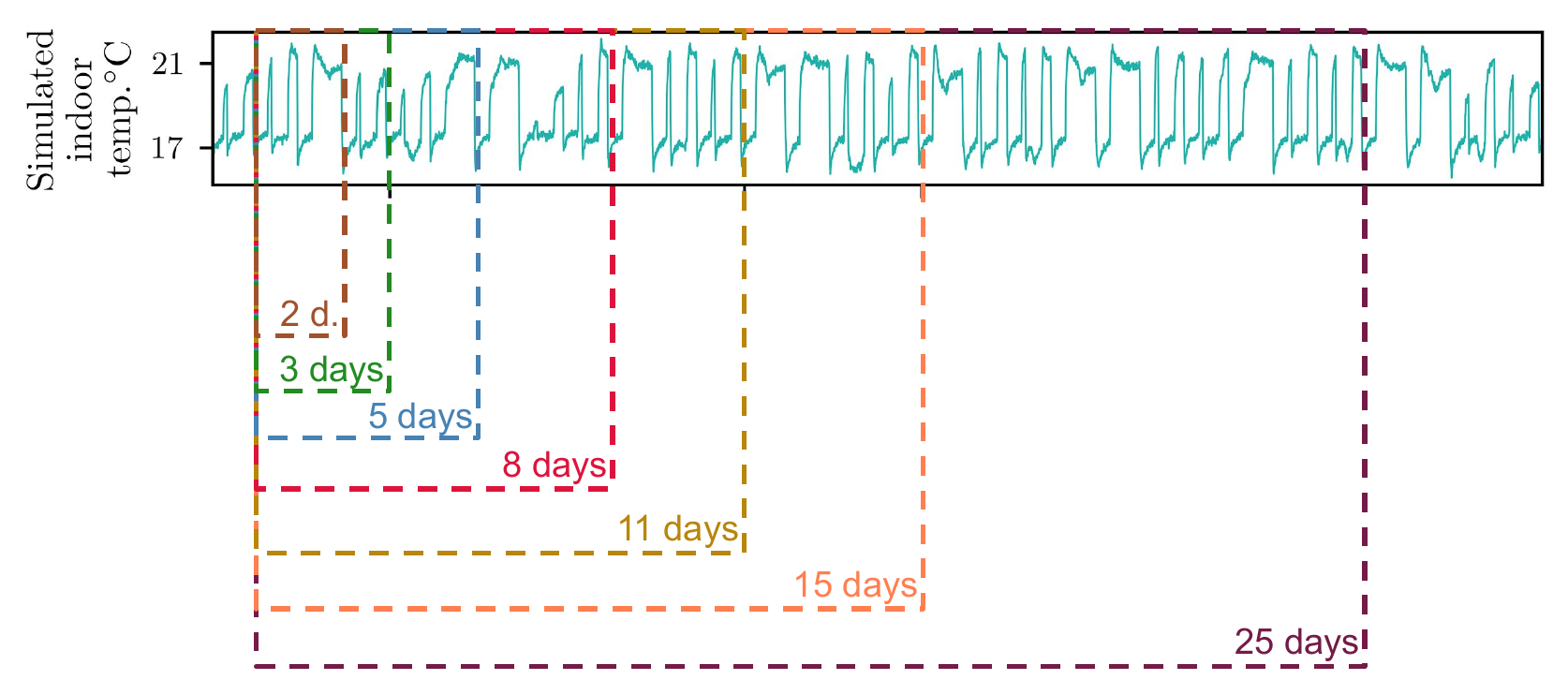}
	\caption{Selection of the data subsets from the stochastic data: 2, 3, 5, 8, 11 and 15 days starting January 2nd}
	\label{fig:frise_temporelle}
\end{figure}

Aiming at a comprehensive comparison, the studied model is calibrated on several subsets of each dataset: 2, 3, 5, 8, 11, 15 days. As shown in figure \ref{fig:frise_temporelle}, all subsets start on January 2nd, i.e. far from the warm up period of Energy Plus, which might have affected the realism of the data. January 2nd is also at the beginning of the month of interest, representative of winter conditions.

\FloatBarrier

\subsection{Step (II): $R_{eq}$ inference}
\label{S:step II}

Inferring a physical property from data means solving an inverse problem. To do so, as detailed in this section, an appropriate model is chosen and its parameters estimated so that the model prediction fits the data. The physical property of interest, here the overall thermal resistance $R_{eq}$ is inferred from the estimated model parameters.

\subsubsection{The choice of stochastic RC models}

\begin{sloppypar}
	The limited number of sensors encourages to consider grey-box models as a data analysis tool. Indeed, as detailed in \cite{Foucquier2013,Li2014}, there are three levels of building thermal modelling: white, grey and black box modelling. Comprehensive thermal dynamic models, i.e. white box models, rely on an extensive number of parameters as they accurately describe the physical behaviour of a building. Therefore, white box models calibrated from poorly informative data will most certainly lead to overfitting and non interpretable parameter values. At the opposite, black box models rely exclusively on statistics and cannot be physically interpreted. In between, grey-box models are a combination of both with physically inspired mathematical structures and statistical modelling to achieve reliable simulation results \cite{Madsen2010,Bohlin1995}. In particular, grey-box models can use a naive description of the building physics to limit the number of parameters \cite{Brastein2019} and still achieve satisfactory modelling through a stochastic diffusion term in the model. Grey box models offer therefore a good compromise in a non intrusive framework.
	
	The numerical procedure is applied to a lumped capacitance model: $T_wT_i$~$R_oR_i$~$A_w$. It is a low order model in the sense that it relies on a system of 2 differential equations \replaced{. It incorporates two thermal resistance parameters $R_o$ and $R_i$, two thermal capacitance parameters $C_w$ and $C_i$ and}{and it incorporates} a single coefficient for solar aperture hereafter named \replaced{$A_w$}{$gA$}. \added{RC models are indeed simplified lumped models of the otherwise non linear thermal exchanges of the building envelope and have physical meaning: thermal capacitances or thermal resistances can be proven to be the lumped capacitances or respectively lumped resistances of each layer of the envelope \cite{Kramer2012, Fraisse2002}.}
	
	As any simplified model, RC state-space models have an intrinsic model error, that can be taken into account as an ad hoc term in the model formulation \cite{Kennedy2001}. If not, Brynjarsdottir and O'Hagan \cite{Brynjarsdottir2014} showed that disregarding model discrepancy may lead to biased and over-confident parameter estimation. Therefore, stochastic differential equations \cite{Madsen1995} are chosen to formulate the RC model, as described in Equation \ref{eq:mod2R2C}.
\end{sloppypar}

\begin{equation}
\label{eq:mod2R2C}
\left\{
\begin{array}{ c c l }
\begin{bmatrix}
\dot{T_w} \\
\dot{T_{in}} 
\end{bmatrix}
&
=
&
\begin{bmatrix}
- \frac{1}{C_wR_o} & \frac{1}{C_wR_i} \\
\frac{1}{C_iR_i} & - \frac{1}{C_iR_i} 
\end{bmatrix}
\begin{bmatrix}
T_w\\
T_{in}
\end{bmatrix}
+
\begin{bmatrix}
\frac{1}{C_wR_o} & \frac{A_w}{C_w} & 0 \\
0 & 0 & \frac{1}{C_i}
\end{bmatrix}
\begin{bmatrix}
T_{out}\\
I\textsubscript{sol}\\
P_h
\end{bmatrix}
+
\sigma
d
\omega
\\
y
&
=
&
\begin{bmatrix}
0 &	1
\end{bmatrix}
\begin{bmatrix}
T_w\\
T_{in}
\end{bmatrix}
+ \epsilon
\end{array}
\right.
\end{equation}

Before all, the structural identifiability needs to be proven, meaning that in theory, given hypothetical ideally informative data, calibrating the model will result into a unique solution. The structural global identifiability of the model is derived by a differential algebra algorithm \cite{Audoly2001} implemented by Bellu et al \cite{Bellu2007} in the tool DAISY.

From the estimation of the \replaced{thermal resistance parameters $R_o$ and $R_i$}{resistive parameters} of the RC model\deleted{s}, $R_{eq}$ can be derived as shown in Equation \ref{eq:calcul_Req}. Equivalent standard deviations are obtained from Equation \ref{eq:calcul std Req}. For readability, the $R_{eq}$ estimations will be given in K/kW which is equivalent to the order $10^3~K/W$.

\begin{equation}
R_{eq} = R_o + R_i
\label{eq:calcul_Req}
\end{equation}

\begin{equation}
\sigma_{R_{eq}} = \sqrt{\sigma_{R_o}^2 + \sigma_{R_i}^2 + 2 \times \sigma_{R_o}\sigma_{R_i}}
\label{eq:calcul std Req}
\end{equation}

\subsubsection{Model calibration}
\label{S:model calibration}
\begin{sloppypar}
	To infer an estimation of the overall thermal resistance, the parameters of the model of interest need to be estimated such as to fit the data. This calibration process is performed by a quasi Newton optimization using the BFGS algorithm, operated in the pySIP python library \cite{Raillon2019}. In particular, the BFGS algorithm minimizes the negative log-likelihood of the model prediction which resumes in finding the most likely set of parameters that fit the selected dataset.
	
	The model calibration is performed in a frequentist approach, where all information about the parameters is supposed to be acquired in the collected data and where the estimation is supposed to have a Gaussian error. Another option, yet more computationally costly, would be to use a Bayesian approach where prior information about the parameter values are incorporated in the calibration process as a prior probability density. However, this paper investigates how weather natural variability influences the amount of information in the data itself and how that reflects on the accuracy and robustness of the estimation of $R_{eqs}$. From this perspective, a frequentist approach makes perfect sense.
	
	As a result, the estimation by the BFGS optimization provides maximum-likelihood (ML-) estimates of the parameters of the selected RC model, with a Gaussian error uncertainty. Both ML-estimates and their uncertainty need to be considered when assessing the outcome of an estimation.
\end{sloppypar}

\subsubsection{Model selection and validation}

Good practice in model calibration demands that model selection be made as to infer results from one best fitting the data. The best fitting model might be different between short and long duration datasets. For this reason, even if model selection is performed on a 3 day dataset, the residuals of prediction of a 15 day dataset are verified as well, to ensure the selected model still performs well on larger datasets.

\begin{figure}
	\centering
	\subfloat[Non filtered prediction on a 3 day dataset]{
		\includegraphics[width=0.7\textwidth]{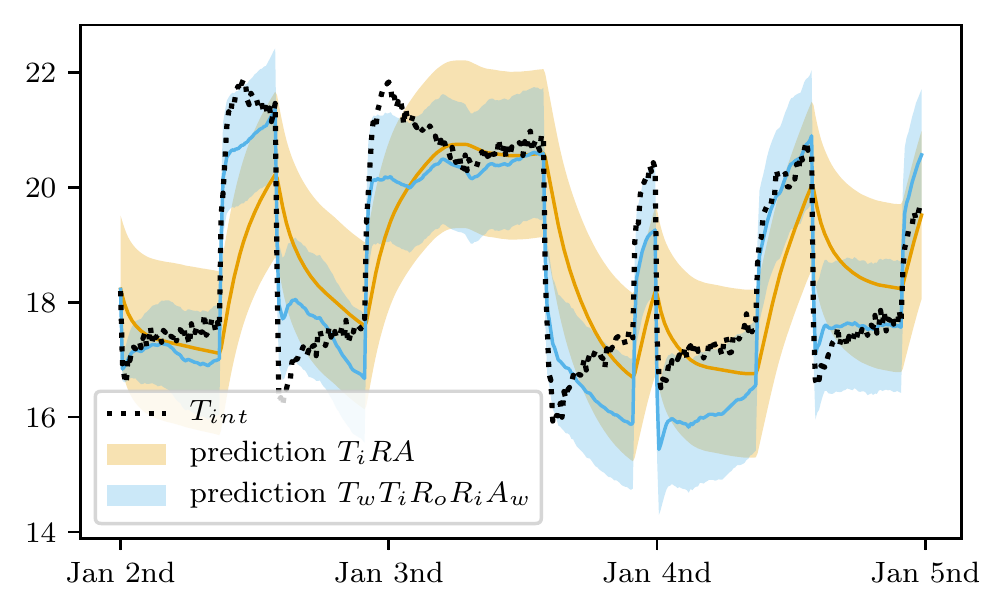}
		\label{fig:pred perf 1st and 2nd}
		}
	\\
	\subfloat[Autocorrelation of the residuals of filtered predictions of models $T_i$ $RA$ and $T_wT_i$ $R_oR_i$ $A_w$ for 3 days data. Basically,this test compares the similarity of the residuals with itself modulo a certain delay, i.e. a lag. Model $T_i$~$RA$ show a significant autocorrelation of its residuals, which implies systematic errors in the prediction. Significant physical phenomena are missing in the model.]{
		\includegraphics[width=\textwidth]{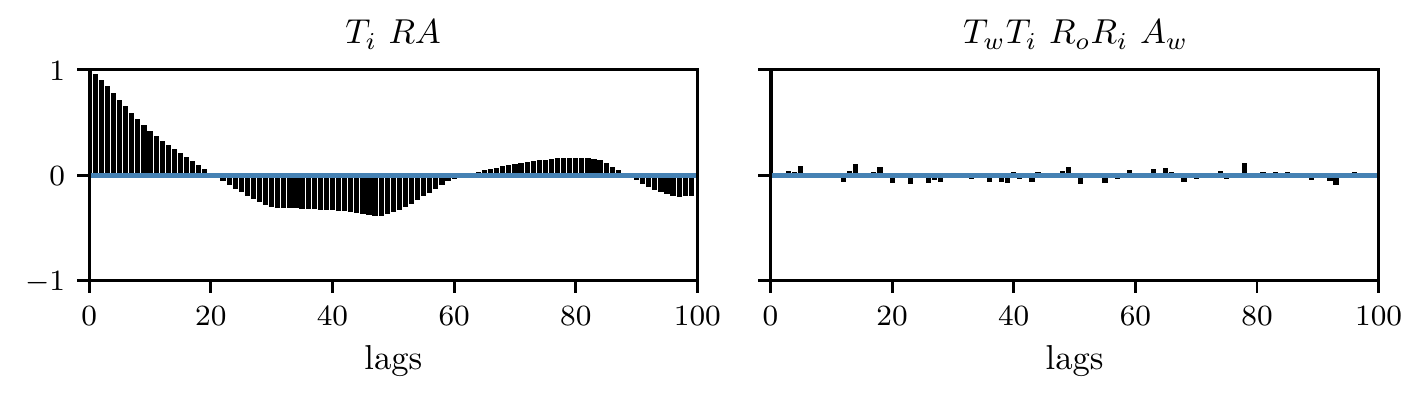}
		\label{fig:autocorr resid 1st and 2nd}
	}
	\caption{Graphical visualisation of how models $T_i$ $RA$ and $T_wT_i$ $R_oR_i$ $A_w$ fit the data}
\end{figure}

First order RC model may be quickly discarded as they visibly fail to catch the physics compared to a second order model, as can be seen in Figure \ref{fig:pred perf 1st and 2nd}. The residuals of a first order model are highly auto-correlated, see Figure \ref{fig:autocorr resid 1st and 2nd}, which again is proof that such a model does not correctly fit the data. A second order model's residuals are indeed much closer to white noise for 3 day calibration. When a 15 day calibration is performed, residuals of prediction show that the first order model is still highly auto-correlated and that the second order model still performs well. Higher order models achieved highly correlated parameter estimations, which is highly undesirable when the parameters need to be physically interpreted.

\subsection{Interpretability: a novel indicator to assess the $R_{eq}$ estimations (step III)}
\label{S:interpret indic}

This section proposes a novel scalar indicator, called the interpretability indicator, to assess the closeness of the estimation to the target value.

\begin{figure}
	\centering
	\includegraphics[width=0.7\textwidth]{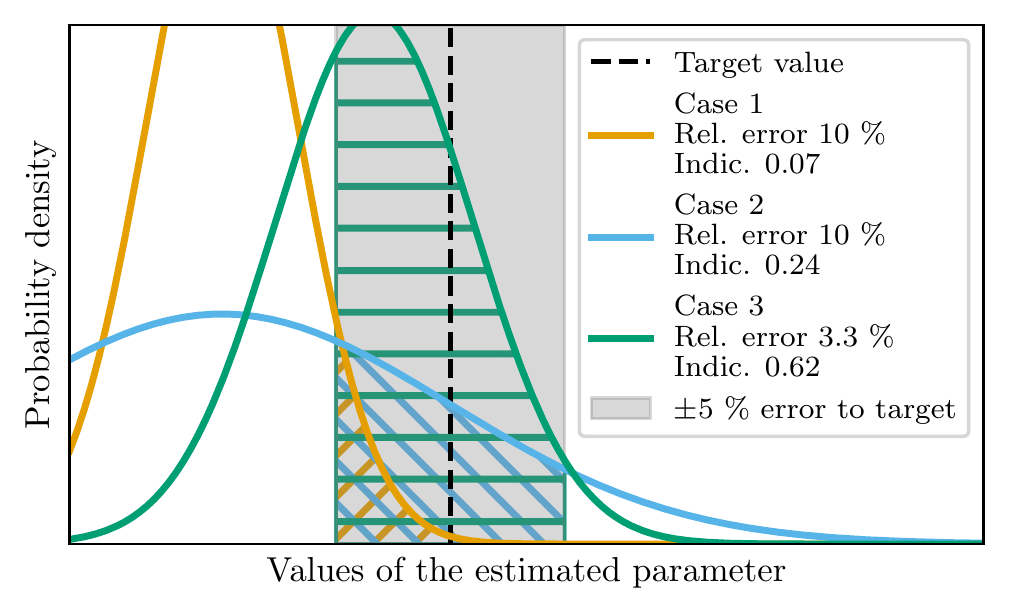}
	\caption{Relative error and interpretability indicator as two accuracy indicators: the relative error reflects poorly on the importance of uncertainty whereas the interpretability indicator accounts simultaneously for accuracy and uncertainty.}
	\label{fig:accuracy indicators}
\end{figure}

\begin{sloppypar}
	The need for a novel indicator comes from the observation that a straightforward relative error of the ML-estimator to the target value is not representative of the uncertainty of such estimation. A relative error cannot indeed discriminate between accurate and uncertain ML-estimations. For example, in Figure \ref{fig:accuracy indicators}, the case 1 estimation is less desirable than the case 2 estimation: both are equally inaccurate from a relative error point of view, but the case 2 reflects appropriately on the inaccuracy through a large confidence interval.
\end{sloppypar}

In order to better discriminate the less desirable estimations from the others, an interpretability indicator is proposed. This indicator represents the area under the bell curve that is $\pm 5$~\% of the target $R^*_{eq}$. The interpretability indicator takes therefore values between 0 and 1. For example in Figure \ref{fig:accuracy indicators}, the case 1 estimation scores close to 0 whereas the case 2 estimation scores at 0.24. Estimations could be considered satisfactory when they score above 0.5, such as case 3 estimation in Figure \ref{fig:accuracy indicators}.

\section{Results}

\subsection{Decreasing variability of the $R_{eq}$ estimation with experiment duration}
\label{S:variability}

\begin{sloppypar}
	As described in the previous section, the assessment methodology has been applied to generate 2000 simulations from 2000 different weather datasets. Each simulation provides 7 synthetic datasets of growing duration used for the calibration of model $T_wT_i$~$R_oR_i$~$A_w$. This sections studies in \ref{S:var 2 day} how the weather natural variability influences a 2-day calibration and establishes in \ref{S:minimal duration} a minimal duration for a robust estimation of $R_{eq}$.
\end{sloppypar}

\subsubsection{Variability with a 2-day calibration}
\label{S:var 2 day}

\begin{figure}
	\centering
	\includegraphics[width=\textwidth]{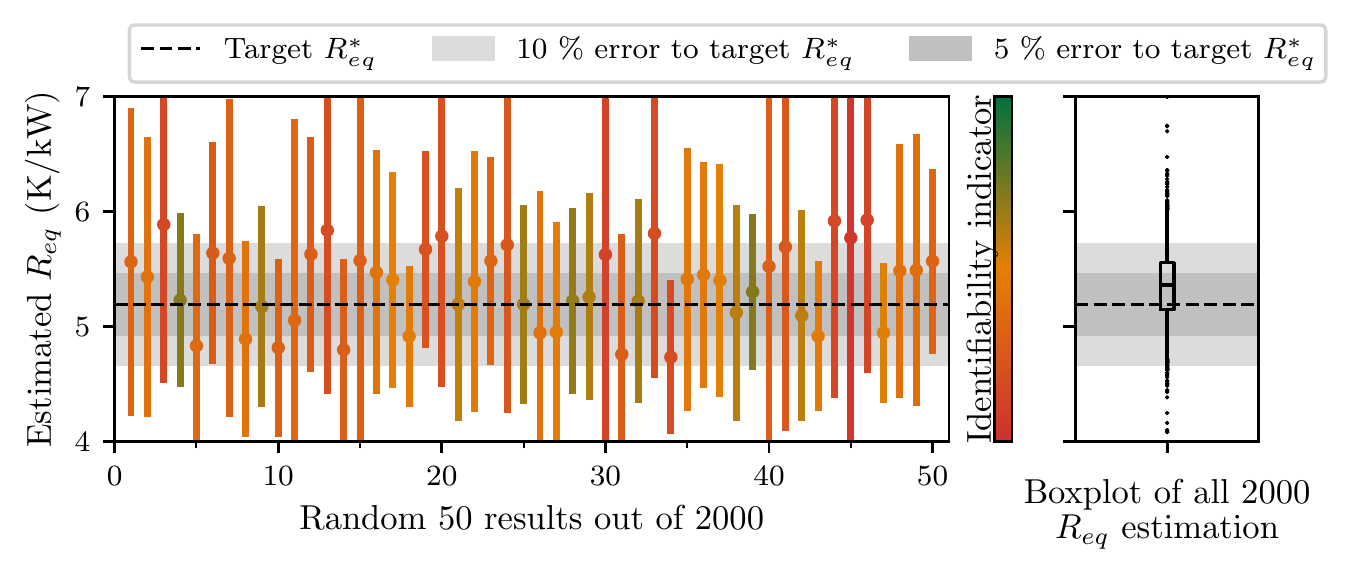}
	\caption{Variability of the $R^*_{eq}$ estimation with 2-days calibration}
	\label{fig:variability 2days}
\end{figure}

For each of the 2000 data sets and for each subset, the stochastic RC model $T_wT_i$ $R_oR_i$ $A_w$ is calibrated. In each case, $R_{eq}$ is inferred as the sum of the resistive parameters estimations. Figure \ref{fig:variability 2days} shows on the left hand side 50 randomly picked maximum likelihood (ML) estimates of $R_{eq}$ with their confidence interval, in order to illustrate the variability of the estimations. The estimations are coloured according to the previously defined interpretability indicator. As a short reminder, it takes values between 0 and 1: the closer to 1, the greener and the closer the estimation to the target value.

\begin{sloppypar}
	Looking at these individual results, there are three cases to distinguish:
	
	\begin{itemize}
		\item the $R_{eq}$ estimation is close to the target $R^*_{eq}$ value: the estimation is accurate and the confidence interval includes the target $R_{eq}$. This case, visible in the greener colours, is the most desirable case;
		\item the $R_{eq}$ estimation is far from the target $R^*_{eq}$ value but the confidence interval includes the target $R^*_{eq}$ or the ML-estimate is accurate with however very large uncertainty: the estimation is not accurate but the credible interval relates to this inaccuracy which keeps the result trustworthy. This case is rendered in orange.
		\item the $R_{eq}$ estimation is far from the target $R^*_{eq}$ value and the confidence interval does not include the target $R^*_{eq}$: not only is the result inaccurate but also give a false sense of confidence on an inaccurate result. These results are visible in red.
	\end{itemize}

	The latter case is the least desirable one but occurs with many estimations. The right hand side of Figure \ref{fig:variability 2days} displays a boxplot af all $R_{eq}$ ML-estimates (dots only) and shows a wide variability: the median of the 2000 ML-estimates falls at $5.36$~$K/kW$ with a standard deviation of $0.35$~$K/kW$ ($5^{th}$ quantile 4.82~$K/kW$ and $95^{th}$ quantile 5.98~$K/kW$). The outlier estimates show absolute errors beyond 20~\% of target $R^*_{eq}$. This variability confirms the preliminary outcomes by suggesting that the influence of weather conditions on the ML-estimates of $R_{eq}$ is not negligible. A data subset longer than 2 days is certainly needed to decrease this variability.
\end{sloppypar}

\subsubsection{Minimal measurement duration for robust model calibration}
\label{S:minimal duration}

\begin{figure}[h!]
	\centering
	\subfloat[2000 $R_{eq}$ ML-estimates for growing duration datasets: datasets over 11 days are all within $\pm 10$~\% error to the target $R^*_{eq}$. Total variances indeed decrease with longer calibration datasets.]{
		\includegraphics[width=\textwidth]{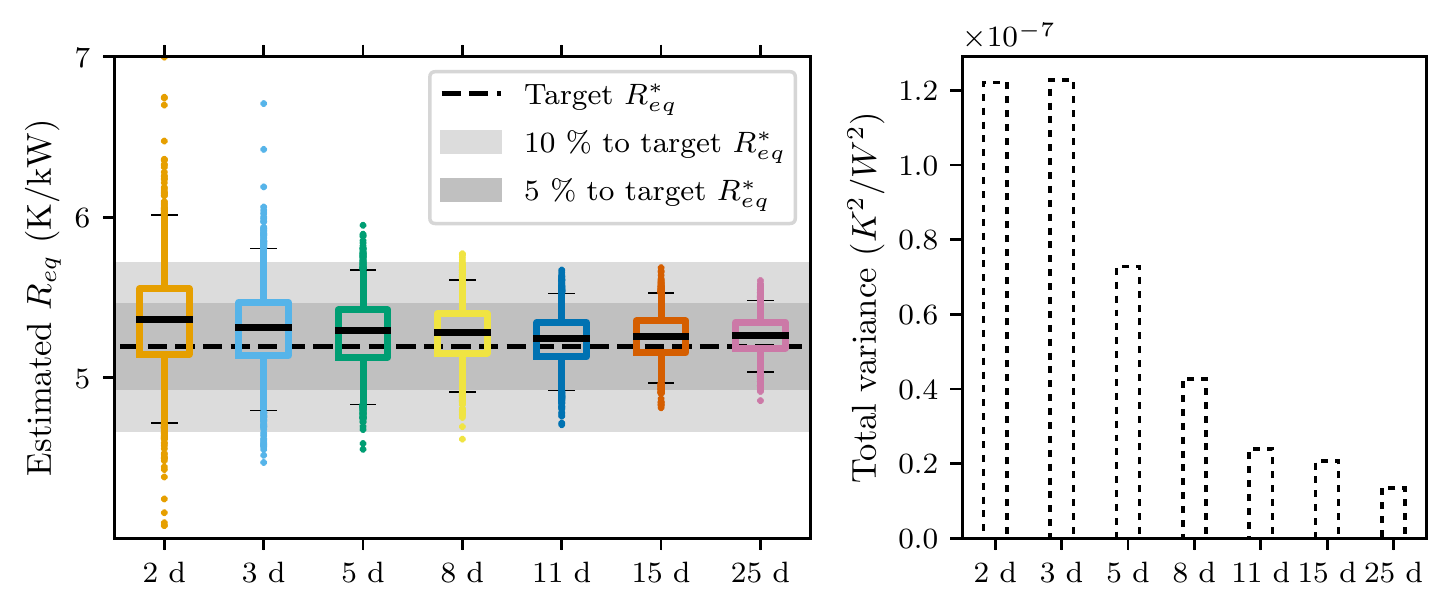}
		\label{fig:evolution variability and variance}
	}
	\\
	\subfloat[Evolution of the 2000 interpretability indicators with growing measurement duration: 80~\% of the 8~day estimations score higher than 0.5, 90~\% of the 11~day estimations and 95~\%of the 25~day estimations.]{
		\includegraphics[width=0.8\textwidth]{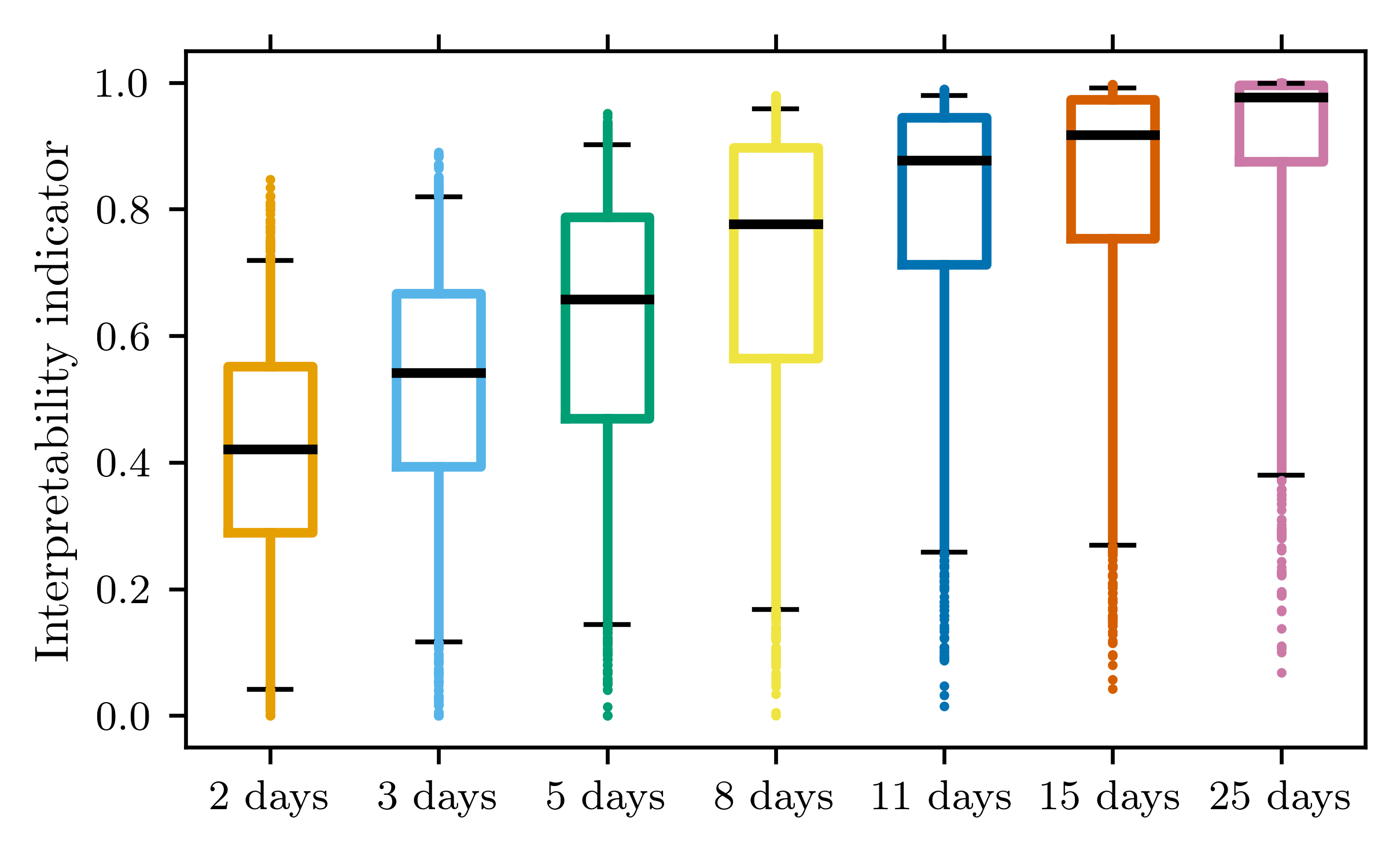}
		\label{fig:evolution indic interpret}
	}
	\caption{Assessing the quality of the $R_{eq}$ estimations through convergence of the ML-estimates within a 10~\% error band and through the interpretability indicator.}
	\label{fig:evolution quality ml estimates}
\end{figure}

The results so far suggest that more than 2 days are necessary to achieve a robust $R_{eq}$ estimation. Figure \ref{fig:evolution variability and variance} shows how the variability of all 2000 $R_{eq}$ ML-estimates varies with the 7 data subsets: model calibration from 2, 3, 5, 8, 11, 15 and 25 day data. From the figure can be inferred that the longer the data subset, the lower the variability. There is distinctively a decrease in total variance towards a median value slightly above the target value $R^*_{eq}$. Calibrations from 11 day data and more show all estimated $R_{eq}$ values within 10~\% of their median value, hence ensures low variability in the $R_{eq}$ estimation with respect to weather influence.

To validate the impression of decrease in variability \replaced[comment=typo]{on}{from} the left \replaced[comment=typo]{hand side }{handside }in Figure \ref{fig:evolution variability and variance}, the right hand side shows for each data subset the evolution of the total variance of ML-estimates. From the 2 day and 3 day data subsets, the $R_{eq}$ ML-estimates have a total variance around $1.2 \times 10^{-7}$~$K^2/W^2$. With 8, 11 and 25 day data subsets, the total variance decreases respectively by a factor 3, 4 and 6. The evolution of the partial variances will be further discussed in the next section.

Studying the variability and variance of the ML-estimates does however not reflect properly on the accuracy of the estimation when considering its uncertainty. Figure \ref{fig:evolution indic interpret} shows therefore the evolution of the interpretability indicator described in section \ref{S:interpret indic} with growing datasets. Longer datasets provide increasing interpretability indicator scores. Considering the minimal score of 0.5 as satisfactory, 80~\% of the 8 day estimations score higher than 0.5, 90~\% of the 11 day estimations and 95~\% of the 25 day estimations. Estimations from 8 day and more datasets can therefore be considered in overall as accurate with a low uncertainty.


As a partial conclusion, 11 day datasets suffice to reduce the error below the $\pm 10$~\% and provide in 90~\% of all cases an acceptable interpretability score. Longer datasets still significantly reduce the overall variance. However, from a practitioner's point of view, longer experiments might be unnecessary, as it would immobilize the experimental setup almost twice as long for an all in all relative decrease in uncertainty.

\FloatBarrier

\subsection{Influential weather variables on an $R_{eq}$ estimation}
\label{S:influential}

\begin{figure}
	\centering
	\subfloat[First order sensitivity indices of the $R_{eq}$ ML-estimates: outdoor temperature and wind speed are the main factors to variability]{
		\includegraphics[width=\textwidth]{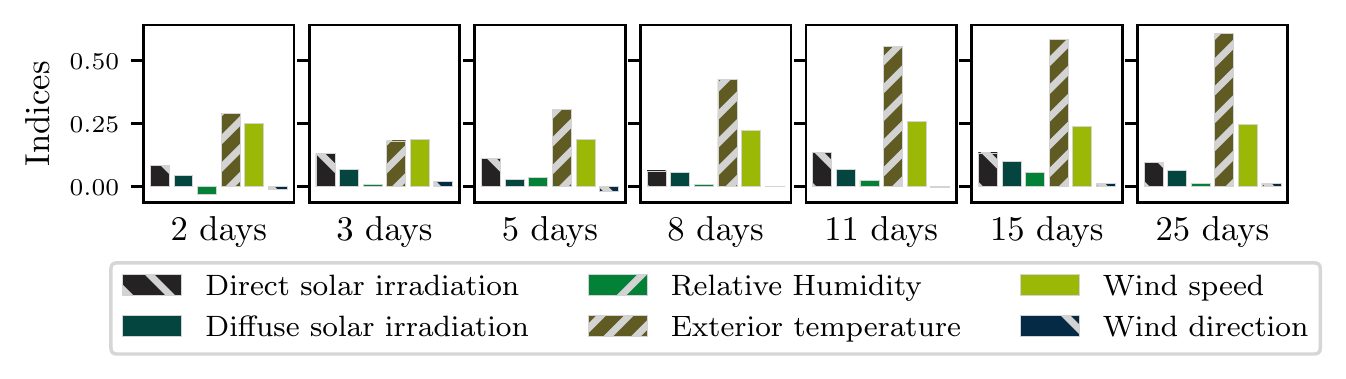}
		\label{fig:SA Req}
	}
	\\
	\subfloat[Decrease of the total and partial variances with longer calibration datasets]{
		\includegraphics[width=0.8\textwidth]{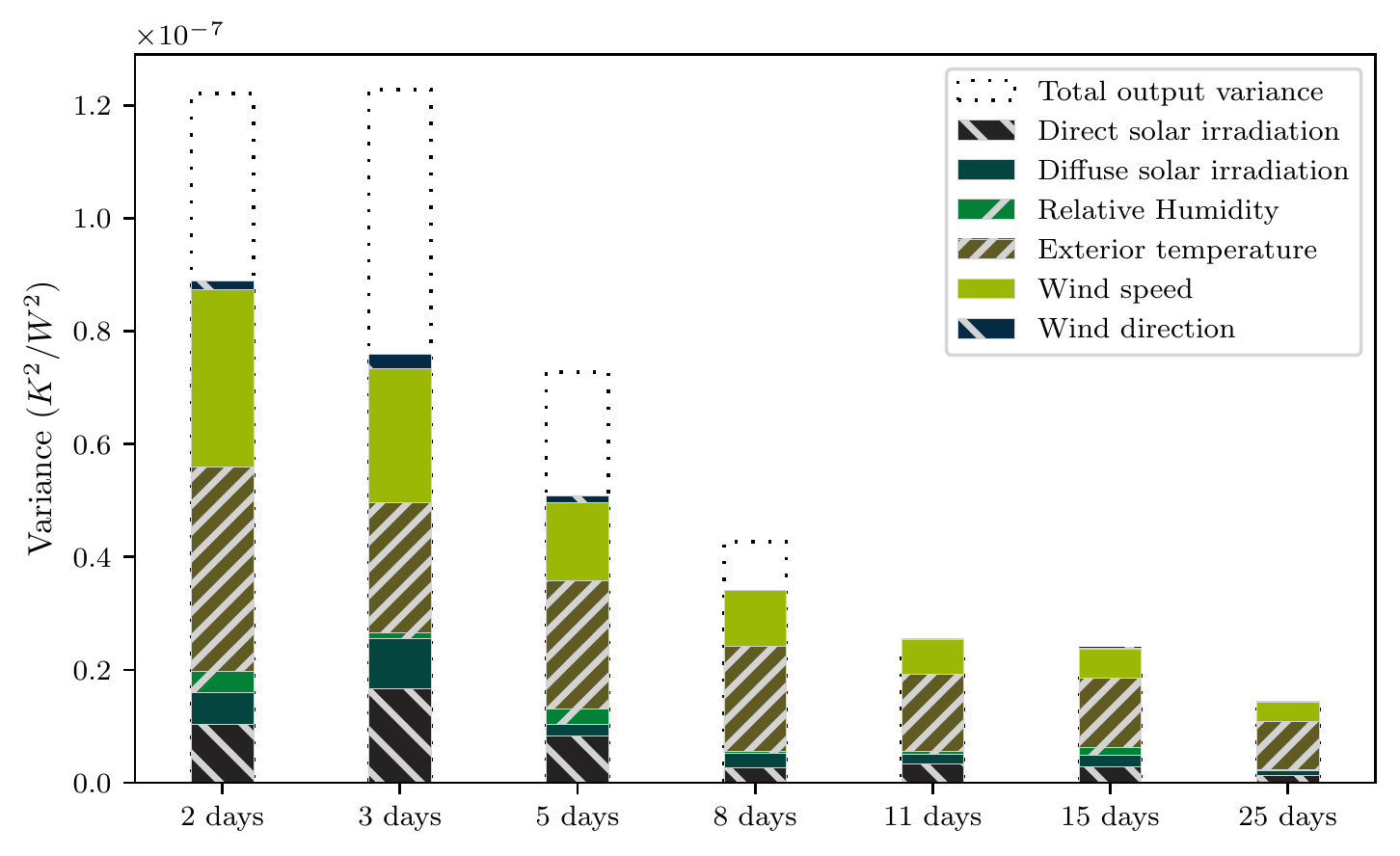}
		\label{fig:evolution partial variance}
	}
	\caption{Individual effect of weather inputs on the estimations of $R_{eq}$}
	\label{fig:indices et variances}
\end{figure}

As mentioned in section \ref{S:stochastic weather}, the synthetic weather files allow a global sensitivity analysis with respect to 6 weather variables. The variability of the $R_{eq}$ ML-estimates can be attributed to the natural variability of these weather variables. Figure \ref{fig:SA Req} shows the sensitivity indices of the estimations of some parameters with respect to the weather variables: $R_{eq}$. The sensitivity indices are calculated for all 7 data subsets. The indices shown in Figure \ref{fig:SA Req} are the first order indices, meaning that they only show the direct influence of each weather variable. If the sum of each first order indices is close to 1, it would imply that there were almost no second order effects, i.e. combined effects of the weather variables. Let us also finally remind that values of sensitivity indices are always simply estimated. The indices given in Figure \ref{fig:SA Req} should mainly be interpreted as order of magnitudes. Indices below $0.1$ may be considered insignificant, given the uncertainty of their estimation.

From Figure \ref{fig:SA Req}, it can be seen that the variability of the $R_{eq}$ ML-estimates is mainly influenced by the outdoor temperature and the wind speed. With shorter datasets, the sum of the first order indices is significantly inferior to 1. This means that the variability is also explained by interactions of weather variables. Variability from with longer datasets is on the contrary almost only explained by the variability of outdoor temperature and wind speed, seeing that the indices add up to 1. Let us also note that neither the relative humidity nor the wind direction were \deleted[comment={gram error}]{not} expected to have an influence on the estimations as \replaced[remark={gram error}]{they are }{it is }not used in the infiltration and ventilation model of EnergyPlus. Their sensitivity indices are indeed insignificant.

Influence of outdoor temperature and wind speed is also visible in the evolution of the partial variances of each weather variable shown in Figure \ref{fig:evolution partial variance}. Let us remind that the total variance is the sum of all order partial variances. The figure shows the first order partial variances, i.e. the partial variances due to the effect of each weather variable individually. With these elements in mind, it is visible that the total variance of the 11, 15 and 25 days datasets is solely explained by first order effects of the weather variables, mainly outdoor temperature and wind speed.

\begin{figure}[h!]
		\centering
		\includegraphics[width=\textwidth]{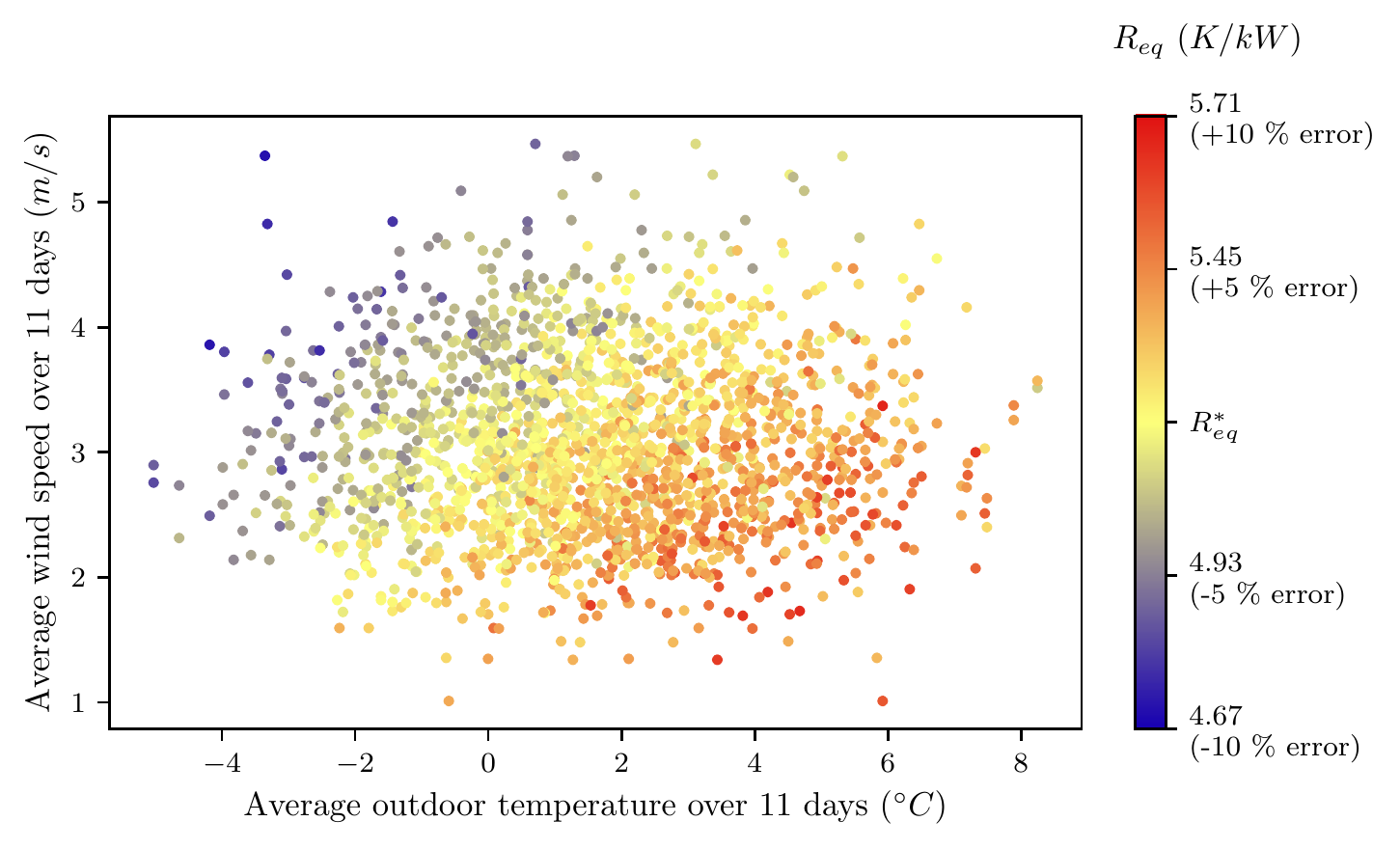}
		\caption{Variability of the $R_{eq}$ ML-estimates from 11 day calibration with respect to outdoor temperature and wind speed. Colours refer to $\pm 10$~\% errors to target $R^*_{eq}$.}
		\label{fig:Influence Temp Wind 11 days}
	\end{figure}

\begin{sloppypar}
	Let us now examine how outdoor temperature and wind speed influence the $R_{eq}$ ML-estimates. Figure \ref{fig:Influence Temp Wind 11 days} shows how the $R_{eq}$ ML-estimates from 11 day datasets vary with the average outdoor temperature on the abscissa and the average wind speed on the ordinate. Warmer periods tend to over-estimate and colder days to under-estimate. At the same time, non windy days produce in overall over-estimations, windy days under-estimations.
	
	At the same time, an interaction can also be seen in Figure \ref{fig:Influence Temp Wind 11 days}: calibration from warm and unwindy days results in over-estimation, cool and windy days in under-estimations. 
	
	This outcome is in agreement with the hypothesis that the large air change rates in the reference model are a cause of inaccuracy in the estimation of the overall thermal resistance. As the ventilation related heat losses have been modelled in the EnergyPlus simulation environment, there is a direct relationship between temperature difference between indoors and outdoors and wind speed. Ventilation related heat losses are larger with cold outdoor temperatures and with high wind speed and on the contrary smaller with warmer and/or unwindy days. It could therefore be expected that acceptable and robust estimations be achieved in less than 11 days in buildings with lower air change rates.
\end{sloppypar}

\begin{figure}[h!]
	\centering
	\includegraphics[width=\textwidth]{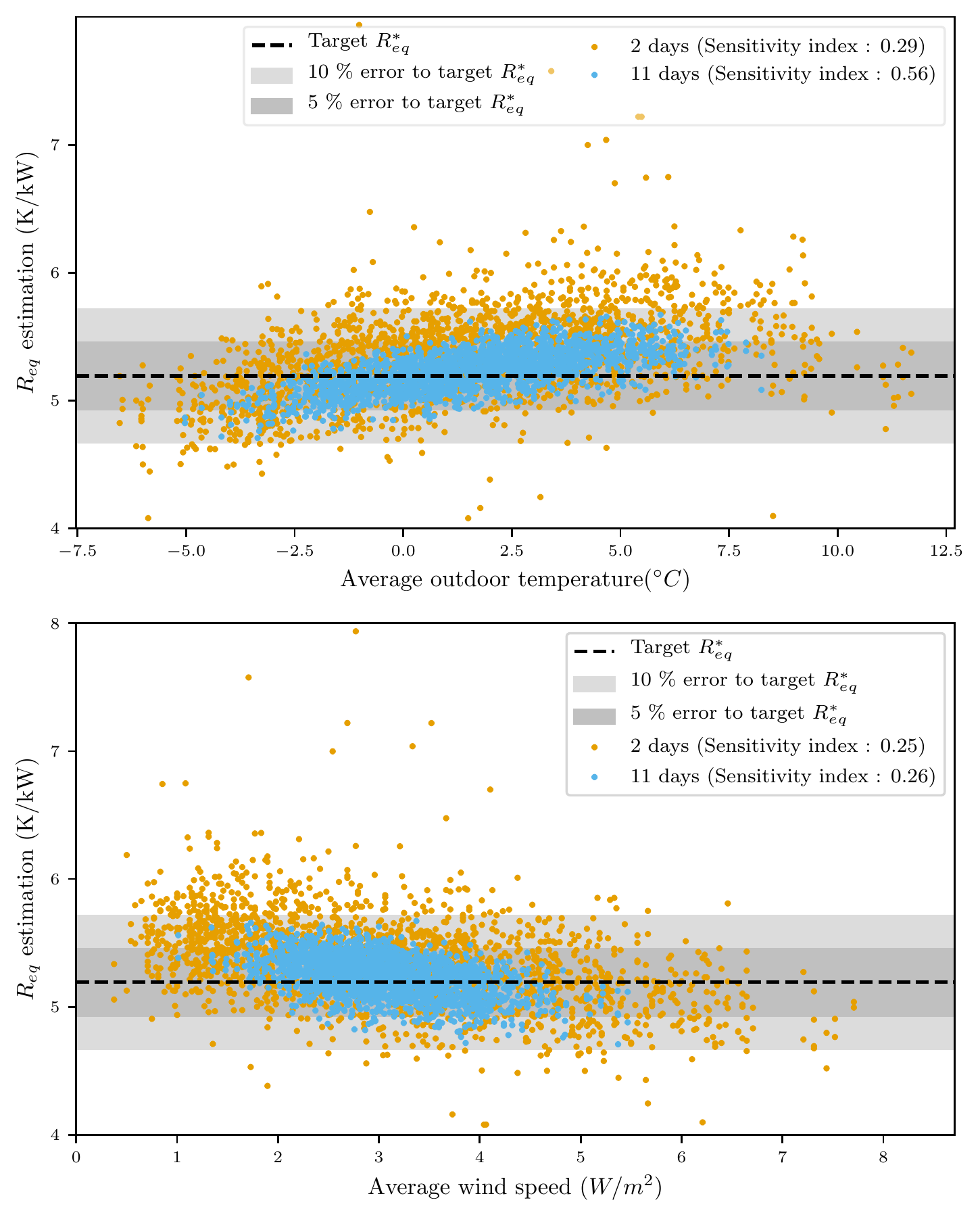}
	\caption{Variability of the $R_{eq}$ ML-estimates with respect to averaged outdoor temperatures and wind speeds, for 2 day and 11 day calibration.}
	\label{fig:Influence Temp Wind 2 11 days}
\end{figure}

\begin{sloppypar}
	Finally, Figure \ref{fig:Influence Temp Wind 2 11 days} shows more clearly how the influence of outdoor temperature and wind speed on the $R_{eq}$ ML-estimates evolves from short to longer datasets.
	
	As seen earlier, this figure too shows the decrease in total variance of the $R_{eq}$ ML-estimates with longer calibration sets: the vertical spread of all estimations are narrower with the 11 day calibration. Interestingly, while the total variability does decrease, the angle representative of the correlation remains quite similar whatever the measurement duration. Longer datasets produce averages that are less spread horizontally, but the relationship between temperature and $R_{eq}$ estimation is almost unaltered.
	
	A natural assumption would have been to consider that colder days lead to more accurate estimations than warmer days, as colder days increase the heat losses and thus the heating power needed to keep up with the indoor temperature set point. This assumption does not seem to hold here. If it were, the variance would be significantly narrower under cold days than under warm days. Here, there is no significant difference in vertical spread between cold and warm days, nor is there any between windy and unwindy days.
\end{sloppypar}

\FloatBarrier

\section{Discussion}
\label{S:Discussion}

\begin{sloppypar}
The results have shown that\added[comment={better highlight the outcomes of this study in the field}]{, in the particular location, climate and season conditions of this study and in the particular case study,} the calibration needs to be based on at least 11 days to ensure convergence within $\pm10$~\% of the target value $R^*_{eq}$. They also showed that the longer the dataset, the smaller the total variance of separate estimations. Calibration from shorter datasets will lead to uncertain results and variability of the estimations will mainly be due to the variability of the weather conditions. Compared to controlled tests with optimized heating or temperature patterns for which a few days is sufficient, 11 days or more is longer. But considered that optimized heating or temperature patterns create richer and less correlated data, it is quite consistent to find 11 days as a minimum in non intrusive conditions.

In addition, as non intrusive measurement design is considered here, 11 days or more is not a prohibitive duration: as long as the test remains user-friendly, leaving data loggers for a few weeks is most probably neither burdensome for the building occupants nor for the expert carrying the diagnosis. All the more, compared to data exploitation at building scale by other low order models such as auto-regressive models as suggested in \cite{Senave2019} or with energy signature methods, \deleted{this }stochastic RC model\added{s} show\deleted{s} to be a faster approach to exploit the data, as long as temperature is not kept constant by the occupants. All day long constant temperatures would necessarily lead to choose models which have the heating power as output to exploit the data.
\\

On another note, the results are certainly specific to this case study\added{. Let us indeed remind that the study has been conducted in winter conditions, concededly typical, and on a particular building type in Geneva which has a temperate oceanic climate \cite{Peel2007}. Whether it is safe to extrapolate the results to other seasons, climates or building types is debatable and needs to be discussed.}

\added[comment={discussion on extrapolation to other climates/locations under the angle of weather natural variability}]{Regarding seasonal and climate related variability of the results, it can be inferred from the results that larger solar irradiations, lower outdoor temperatures or larger wind speeds will affect the outcomes. This might be the case when such experiment is performed in autumn or spring weather, or in colder or more windy climates. Yet, let us also remind that a same building in different locations will have a different target value $R^*_{eq}$ as it includes heat losses by infiltration. In the end, the results will probably slightly change in much colder or more windy locations, but the extrapolation would still be feasible. It could then be expected that}\deleted{although} the order of magnitude of the calibration duration would be approximately similar.\added{ Such extrapolation would however be more risky if the natural variability of one of the weather variables is significantly larger or narrower than that in Geneva. In a colder climate for example, but with little variability in a winter month, lower outdoor temperatures will certainly have an influence on the amount of heat losses through air change in the building, but will also affect the target value through its air change rate component. In the end, regardless of how cold it might be, low natural variability of outdoor temperatures will barely affect the variability of the $R_{eq}$ estimations. For these reasons, the results of this paper can still be viewed as a benchmark in the field of non intrusive measurements exploitation. The 11 day minimal duration can then serve as a comparison for other locations, as long as natural variability of the weather variables is considered.}

\added[comment={why summer conditions are not in the scope of this paper}]{Summer conditions, as well as dry or tropical climates where active cooling is needed, are however absolutely out of the scope of the tested conditions of this paper because the proposed experiment uses a heating power signal as model input. If a cooling power based appropriate methodology were to be developed, the present study may suggest that the variability of the thermal characterisation estimation could be influenced by much larger solar irradiation, correlated to high outdoor temperatures. A minimal measurement duration could not however be safely determined from the outcomes of this study and would need further dedicated investigations.}

\added[comment={validity for other building types}]{Another distinctive feature of this study is the particularity of the case study: a one storey internally insulated house, with high insulation in the attics and under the ground floor. Although already discussed, its large air change rate is also a particularity. Whether the outcomes are valid for building with heated or unheated neighbouring zones, for other levels of insulation or for larger buildings such as apartment blocks remains uncertain, not only because the effect of weather variability on the thermal resistance estimation would be different, but also because these conditions raise questions of the feasibility of such non intrusive experiment in the first place. Further work will be necessary to assert the feasibility of non intrusive thermal characterisation in other building types.\\}

\replaced{A future major development would also be to exploit an actual measurement campaign and for that purpose, an important indicator would actually be to consider convergence of the estimation.}{Upon exploitation of an actual measurement campaign, an important indicator would actually be to consider convergence of the estimation.} \deleted{ }Let us therefore take the opportunity to make a distinction between convergence of a single estimation and repeatability of such an experiment. At the scale of a punctual measurement campaign, the stop factor would be convergence of the estimation results: continuing the measurements does not significantly change the results.

Although tools for assessing the convergence are not the purpose of this paper, one may extrapolate the well-established ISO 9869 standard \cite{ISO9869} criteria for wall-scale characterisation to building scale and for results from RC models. First, the $R_{eq}$ estimation should not deviate more than 5\% from the 24~h earlier estimation. Secondly, with $N$ the total duration of measurements, the $R_{eq}$ value inferred from the last $2/3N$ days are within 5~\% of the first $2/3N$ days.
\end{sloppypar}

\begin{figure}
	\centering
	\includegraphics[width=\textwidth]{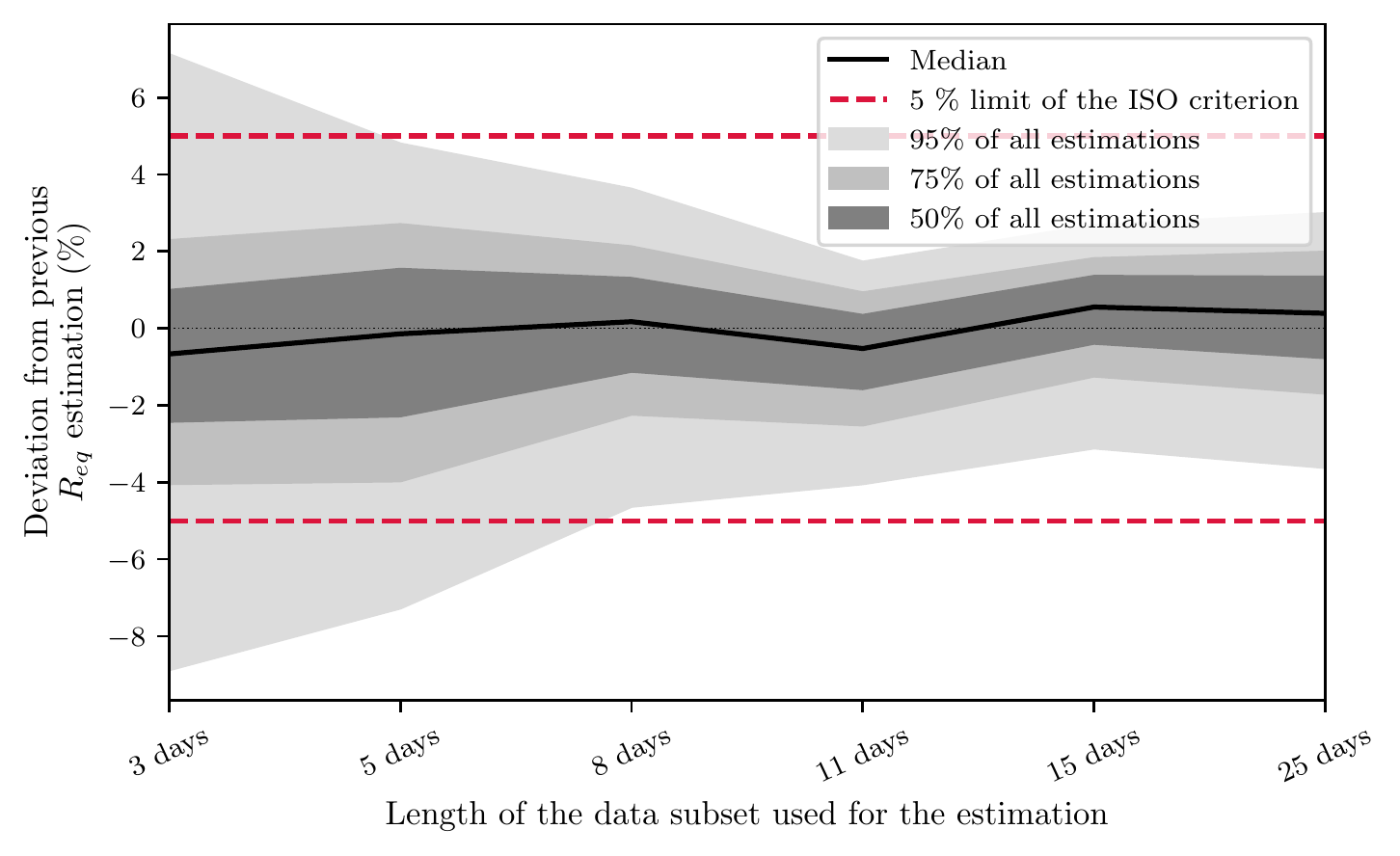}
	\caption{ISO 9869-1 convergence criteria: there should be less than 5~\% deviation with the previous estimation.}
	\label{fig:Convergence ISO}
\end{figure}

The second ISO 9869 standard criterion is roughly applied with the available data in Figure \ref{fig:Convergence ISO}. The estimation from a 3 day dataset is compared with that from a 2 day dataset by calculating a deviation percentage: $\delta_{i} = (R_{eq}^{3 \quad days} - R_{eq}^{2days}) / R_{eq}^{2days} * 100$. Then, the estimation from a 5 day dataset is compared to that from a 3 day dataset and so forth. The calculated deviation from the previous estimation is considered satisfactory when it scores below $5~\%$.

All 2000 deviations calculations are represented as grey shaded areas with 50~\%, 75~\% and 95~\% quantiles. Interestingly, a large majority of cases show a convergence in the sense of the second ISO 9869 standard criteria within 5 days. 8 day datasets are sufficient for convergence in more than 95~\% of all cases. Yet at the same time, repeatability as defined previously is not quite achieved: the variance caused by weather conditions is still significant.

The third criterion is not literally applicable to the data from this paper, but one may extrapolate that there could be up to 10~\% deviation between the first $2/3 \cdot N$ days inference and the last $2/3 \cdot N$ days inference, for example when the first days are particularly cold and the last particularly warm or inversely. However, with similar consecutive weather conditions, convergence would be considered "achieved" rather quickly.

In summary of the convergence topic, according to these criteria, convergence could in some cases be considered as achieved with fewer measurements than the results from this paper would suggest to reach repeatability.

The question is then maybe not to look at deviation in the punctual $R_{eq}$ result but rather quantify the information learnt from the data and its evolution, with in mind the representativeness of the weather conditions. In this light, it is a call for a Bayesian perspective on the results: the important outcome to consider is the complete posterior distribution and not the single most probable value of interest. If upon significant variation of weather there is no more learning from the data, i.e. there is no further change in the posterior distribution, then the measurements may stop.

\begin{figure}
	\centering
	\includegraphics[width=\textwidth]{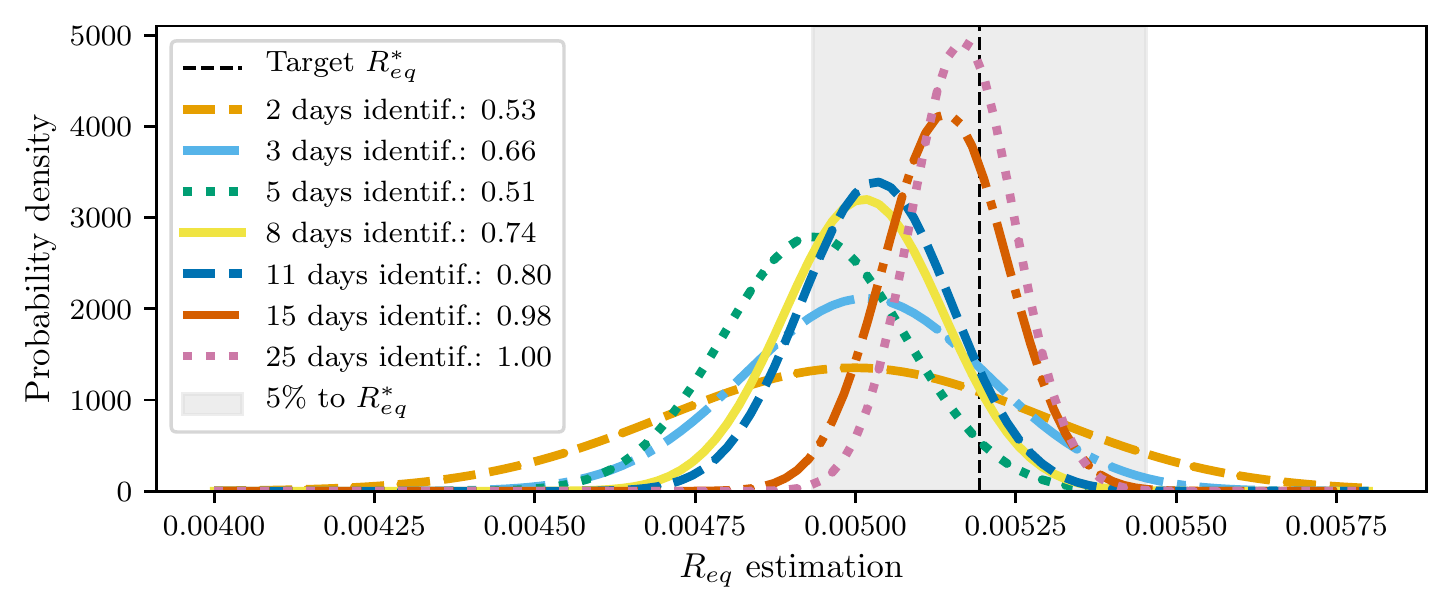}
	\caption{Illustration of the identifiability indicator: the most important is that most of the posterior distribution is satisfactorily close to the target value. Although not perfectly accurate at peak, the posterior distribution is in overall within boundaries of the grey target area, provided sufficient data for calibration.}
	\label{fig:illustration convergence indicator}
\end{figure}

Figure \ref{fig:illustration convergence indicator} illustrates for one case how the posterior distribution of a $R_{eq}$ estimation varies with longer dataset. As a comparison, a $5~\%$ error area around the target value is represented in grey. An estimation from a 2 or 3 day dataset does not provide a satisfactory estimation, seen that the uncertainties are large. This means that the amount of useful information in short datasets is insufficient for trustworthy estimation. A 5 day calibration provides an under-estimation, which is probably related to particular weather conditions. Estimations from 8 or 11 day datasets are rather similar and could be considered satisfactory. This suggests that data from 8 or 11 days is sufficiently rich to provide a$R_{eq}$ estimation. Finally, estimations from 15 or 25 days are very accurate as their interpretability indicator scores at almost $1.0$.

Figure \ref{fig:illustration convergence indicator} illustrates therefore how a posterior distribution gives a wider perspective on the convergence of the estimation than the only ML-estimator as suggested in the ISO 9869 standard \cite{ISO9869}. The interpretability indicator as defined in relation to the target and henceforth known value can obviously not be a metric for convergence. It however supports the idea of judging convergence through posterior distribution, by the use of a divergence metric such as the Kullback Leibler divergence. Furthermore, it would make sense to exploit data in a Bayesian approach and for example use in line calibration algorithms like Sequential Monte Carlo (see \cite{Rouchier2019}).

\FloatBarrier

\section{Conclusion}
\begin{sloppypar}
	Establishing reliable methods for estimating the thermal performance of buildings remains a challenging issue under the constraint of a non intrusive measurement framework: the data is collected in non intrusive conditions,where the indoor air temperature is controlled as to provide occupant-friendly conditions and is thus less informative. Such data may therefore lead to errors in the thermal diagnosis and the outcome may be uncommonly influenced by the boundary conditions, i.e. the weather conditions.
	
	In this context, this paper has developed an original model assessment framework to investigate the influence of weather conditions on the repeatability and hence feasibility of the thermal characterisation of a building envelope from measurements in non intrusive conditions. The methodology relies on a sensitivity and uncertainty analysis of the overall thermal resistance $R_{eq}$ estimation with respect to 6 weather variables.
	
	\added{The proposed methodology has proven to be effective to assess the robustness of the overall thermal resistance estimation. Through the analysis of the variability of all estimations over different measurement durations and through the analysis of their partial variances and sensitivity indices, the minimal measurement duration can be assessed and the main influential weather variables identified.}
	
	\replaced{The paper shows how the methodology is applied to a case study and how a}{A} stochastic RC model is used to exploit the data generated by the model assessment framework. After comparing 2, 3, 5, 8, 11, 15 and 25 days of model calibration, it has been found that 11 days and longer provide repeatable results regardless of the outdoor conditions.
	
	The variability of the \added[comment={avoiding non self standing abbreviations in the conclusion}]{overall thermal resistances }$R_{eq}$ estimations from  11 days and longer observed in the outcomes is in the present case study exclusively due to the variability of outdoor temperature and wind speed. This case study has indeed large air change rates which would emphasize the effect of these two weather \added[remark={typo}]{variables} on the overall heat transfers.
	
	\added[comment={better highlight the importance and validity of this work in the field}]{Although the 11 day duration is strictly speaking specific to the particular climate conditions and the particular case study, the strength of the uncertainty and sensitivity analysis of this methodology allows to prudently extend the validity of the results to other cases, as long as similar weather variability remains. Indeed, the robustness of the overall thermal resistance estimation is not simply studied in relationship to the weather conditions seen as absolute values, but rather to weather variability itself.} \replaced{The}{A} minimal duration of 11 days found in this application gives \added{therefore }a sense of the order of magnitude of duration that can be expected from the exploitation of data with stochastic RC models \added{and may serve as a benchmark for future investigations}. \deleted{Yet, 11 days is certainly strictly speaking specific to the case study used in the model assessment methodology. }\replaced{In addition, t}{T}he results \deleted{therefore} call for further effort on establishing reliable tools for the assessment of convergence from a given dataset to later exploit actual on site experiments.
\end{sloppypar}

\section*{Acknowledgements}

The authors would like to thank the French National
Research Agency (ANR) for funding this work
through the BAYREB research project (ANR-15-
CE22-0003).

\section*{CRediT authorship contribution statement}
\begin{sloppypar}
	\added[comment={Compliance Applied Energy guidelines}]{All authors have equally contributed to the conceptualization and methodology developped in this study. Sarah Juricic and Jeanne Goffart provided the simulated data, with the necessary support of Nicolas Cellier for the numerical implementation. Sarah Juricic conducted the formal analysis, validation and visualisation, with careful supervision of all authors. The original draft was written by Sarah Juricic, with thorough review by all authors. Funding acquisition has been performed by Simon Rouchier.}
\end{sloppypar}

\section*{Declaration of Competing Interest}
\added[comment={Compliance Applied Energy guidelines}]{The authors declare that they have no known competing financial interests or personal relationships that could have appeared to influence the work reported in this paper.}


\bibliography{mybibfile}

\end{document}